\documentclass[%
 reprint,
 aip,
showkeys,
 amsmath,amssymb,
longbibliography,
]{revtex4-1}

\usepackage{graphicx}
\usepackage[caption=false]{subfig} 
\usepackage{pifont}

\usepackage{dcolumn}
\usepackage{bm}

\newcommand{\ee}{\end{equation}} 
\newcommand{\be}{\begin{equation}}

\usepackage[mathscr]{euscript}  
\usepackage{xcolor}  

\usepackage{tcolorbox}
\usepackage{comment}
\usepackage{float}

\makeatletter
\newsavebox{\@brx}
\newcommand{\llangle}[1][]{\savebox{\@brx}{\(\m@th{#1\langle}\)}%
  \mathopen{\copy\@brx\kern-0.5\wd\@brx\usebox{\@brx}}}
\newcommand{\rrangle}[1][]{\savebox{\@brx}{\(\m@th{#1\rangle}\)}%
  \mathclose{\copy\@brx\kern-0.5\wd\@brx\usebox{\@brx}}}
\makeatother

\makeatletter
\newcommand{\vast}{\bBigg@{3.4}}
\makeatother

\usepackage{tikz}

\makeatletter
\newcommand\emailx[1]{%
\move@AF%
\def\@affil{{\normalfont\,#1\strut}{}}%
}%

\begin{document}

\preprint{ApS/123-QED}

\title{Shear-Driven Diffusion with Stochastic Resetting}

\author{Iman Abdoli}
\email{iman.abdoli@hhu.de}
\affiliation{Institut für Theoretische Physik II - Weiche Materie, Heinrich-Heine-Universität Düsseldorf, D-40225 Düsseldorf, Germany}

\author{Kristian St\o{}levik Olsen}
\affiliation{Institut für Theoretische Physik II - Weiche Materie, Heinrich-Heine-Universität Düsseldorf, D-40225 Düsseldorf, Germany}

\author{Hartmut L\"{o}wen}
\affiliation{Institut für Theoretische Physik II - Weiche Materie, Heinrich-Heine-Universität Düsseldorf, D-40225 Düsseldorf, Germany}

\begin{abstract}
External flows, such as shear flow, add directional biases to particle motion, introducing anisotropic behavior into the system.
Here, we explore the non-equilibrium dynamics that emerge from the interplay between linear shear flow and stochastic resetting. The particle diffuses with a constant diffusion coefficient while simultaneously experiencing linear shear and being stochastically returned to its initial position at a constant rate. We perturbatively derive the steady-state probability distribution that captures the effects of shear-induced anisotropy on the spatial structure of the distribution.  We show that the dynamics perform a crossover from a diffusive to a super-diffusive regime, which cease to exist in the absence of shear flow. We also show that the skewness has a non-monotonic behavior when one passes from the shear-dominated to the resetting-dominated regime. We demonstrate that at small resetting rates, while resetting events are rare, they incur a large energetic cost to sustain the non-equilibrium steady state. Surprisingly, if only the $x$-position is reset, the system  never reaches a steady state but instead spreads diffusively.




\end{abstract}

\maketitle

\section{Introduction}
Brownian motion through fluids is of importance to a wide range of scientific fields, with examples including tracers in biological environments, Brownian motion 
under external forces, and the control of colloids in technological applications, among others
~\cite{eckstein1977self, morris1996self, frankel1991generalized, san1979brownian,novikov1958concerning, makuch2020diffusion, abdoli2023tailoring, tothova2024brownian}. In the simplest case, Brownian motion in a fluid at rest gives rise to the well-known diffusive dynamics initially described by Einstein and Smoluchowski, where particles undergo random, thermally-driven motion in an isotropic medium~\cite{einstein1906theorie, von1906kinetischen}. 
However, more complex flow fields and couplings between particles and fluids can produce a rich variety of behaviors, such as 
resonant effects, Taylor-Aris dispersion, and anomalous diffusion~\cite{taylor1953dispersion, aris1956dispersion, belongia1997measurements, huang2011direct, kahlert2012resonant, kumar2021taylor}. Further investigations into active particle dynamics have extended this understanding by exploring the complex interplay between self-propulsion, deformation, and external flow fields, particularly in shear flows~\cite{ten2011brownian, tarama2013dynamics, sandoval2018self, asheichyk2021brownian}.

In simple sheared fluids, flow-induced cross-correlations emerge due to the interaction between the fluid's motion and the particle's random walk. These cross-correlations reflect the particle’s ability to perform stochastic jumps across streamlines, leading to anisotropic diffusion and more complex transport behaviors~\cite{ziehl2009direct}. For example, in confined geometries, such as microfluidic channels or harmonic potentials created by optical tweezers, shear flow can induce directional bias in particle motion, significantly altering the statistics of displacement and velocity fluctuations~\cite{howard2016axial, holzer2010dynamics, bammert2010probability}. Recent experimental advances have explored anomalous diffusion in sheared diffusive systems, further enriching our understanding of tracer dynamics under flow conditions \cite{orihara2011brownian}.
From an experimental perspective, the statistics of a confined particle in a flow field is more compliant than a free particle, the latter of which is more amenable to large displacements and fluctuations. For example, the cross-correlations that emerge in shear flow has been measured by holding the particle in a harmonic trap generated by optical tweezers \cite{ziehl2009direct}.

An alternative, less invasive method of introducing confinement is stochastic resetting, where a particle is allowed to freely explore the sheared fluid but is intermittently returned to a prescribed position at random intervals~\cite{evans2011diffusion, evans2011optimal}. This resetting mechanism disrupts the system's natural evolution
, breaking time-reversal symmetry and leading to complex steady states with novel relaxation properties \cite{evans2020review,majumdar2015dynamical,santra2020run}. Such non-equilibrium dynamics have garnered significant attention across various fields, including statistical physics, search optimization~\cite{kusmierz2015optimal, reuveni2016optimal,evans2024stochasticresettingprevailssharp}, and biological systems~\cite{rotbart2015michaelis}, where resetting can optimize efficiency or regulate system behavior. 




While much is known about resetting in simple diffusive systems, less is understood about how it interacts with more complex environments, particularly those involving external forces or flow fields. Recent developments in this direction includes the study of the combined effect of resetting and spatially or temporally disordered diffusion coefficients \cite{sandev2022heterogeneous,wang2021time,ray2020space,mutothya2021first,lenzi2022transient,sandev2024fractional,menon2024random,sandev2022stochastic,bressloff2020switching},  resetting in external magnetic fields~\cite{abdoli2020stationary, abdoli2021stochastic}, resetting in complex geometries \cite{domazetoski2020stochastic,antonio2020comb,maso2023random,singh2021backbone,trajanovski2023ornstein}, and processes with spatial or temporal modulation of the resetting rate \cite{pal2016diffusion,garcia2023optimal,evans2011optimal,roldan2017path,kusmierz2019robust,pinsky2020diffusive}. Resetting in systems where the dynamics have a bias, such as asymmetric random walks or drift-diffusion processes, has been studied, which could correspond to the simplest case of a time-independent spatially homogeneous fluid flow \cite{sandev2021diffusion,pierce2022advection,ray2019peclet,stanislavsky2022subdiffusive,ahmad2019first,vinod2022time,sarkar2022biased,montero2017continuous}.  


In this work, we investigate the interplay of linear shear flow and stochastic resetting in a two-dimensional diffusive system. The shear flow in the $x$-direction creates a situation where the velocity of a particle depends linearly on its position, leading to a differential movement in the horizontal direction. This introduces an asymmetry in the dynamics, where motion in the vertical direction remains unaffected by shear, while motion in the horizontal direction becomes biased. This asymmetry results in an anisotropic steady-state distribution, which deviates from the isotropic distributions typically found in purely diffusive resetting systems. Using perturbative methods, for small shear rates $\dot\gamma\ll 1$,  we derive an approximate expression for the steady-state probability distribution of the particle's position that captures the anisotropy induced by the shear; the system is characterized by a diffusive core, with a correction term that reflects the influence of the shear flow. This correction term introduces anisotropy into the spatial structure of the distribution, leading to different behaviors in the two spatial directions.

For an arbitrary shear rate, we use the method of moments to exactly compute the first four cumulants of the propagator. 
When resetting to a location to high shear flow, the system exhibits a crossover from diffusive to super-diffusive motion before reaching a steady state. When resetting to regions of lower shear, the super-diffusive regime is no longer observed.  The steady state also gains cross correlations due to the simultaneous presence of resetting and shear flow. The skewness has a non-monotonic behavior when one passes from the shear-dominated (i.e., $\dot\gamma\gg r$) to the resetting-dominated (i.e., $r\gg\dot\gamma$) regime and is zero when the problem is symmetric. 
The kurtosis in the $x$-direction shows high non-Gaussianity in the shear-dominated regime, and saturates to $6$ in the resetting-dominated regime, which is known for the case of one-dimensional diffusion with resetting. 

Furthermore, we show that at small resetting rates, even though a resetting event is rare, once it occurs there will be a large energetic cost to maintain the non-equilibrium steady state resulting from resetting. Finally, we demonstrate that if we only reset the $x$-position of the particle, surprisingly, the system never reaches a steady state but rather spread diffusively. This is in spite of the non-Gaussinanity of the probability density encoded in the kurtosis indicating an example of Brownian yet non-Gaussian diffusion~\cite{chechkin2017brownian, sposini2018random, postnikov2020brownian}. 

\begin{figure}[t]
    \centering
    \includegraphics[width=8cm]{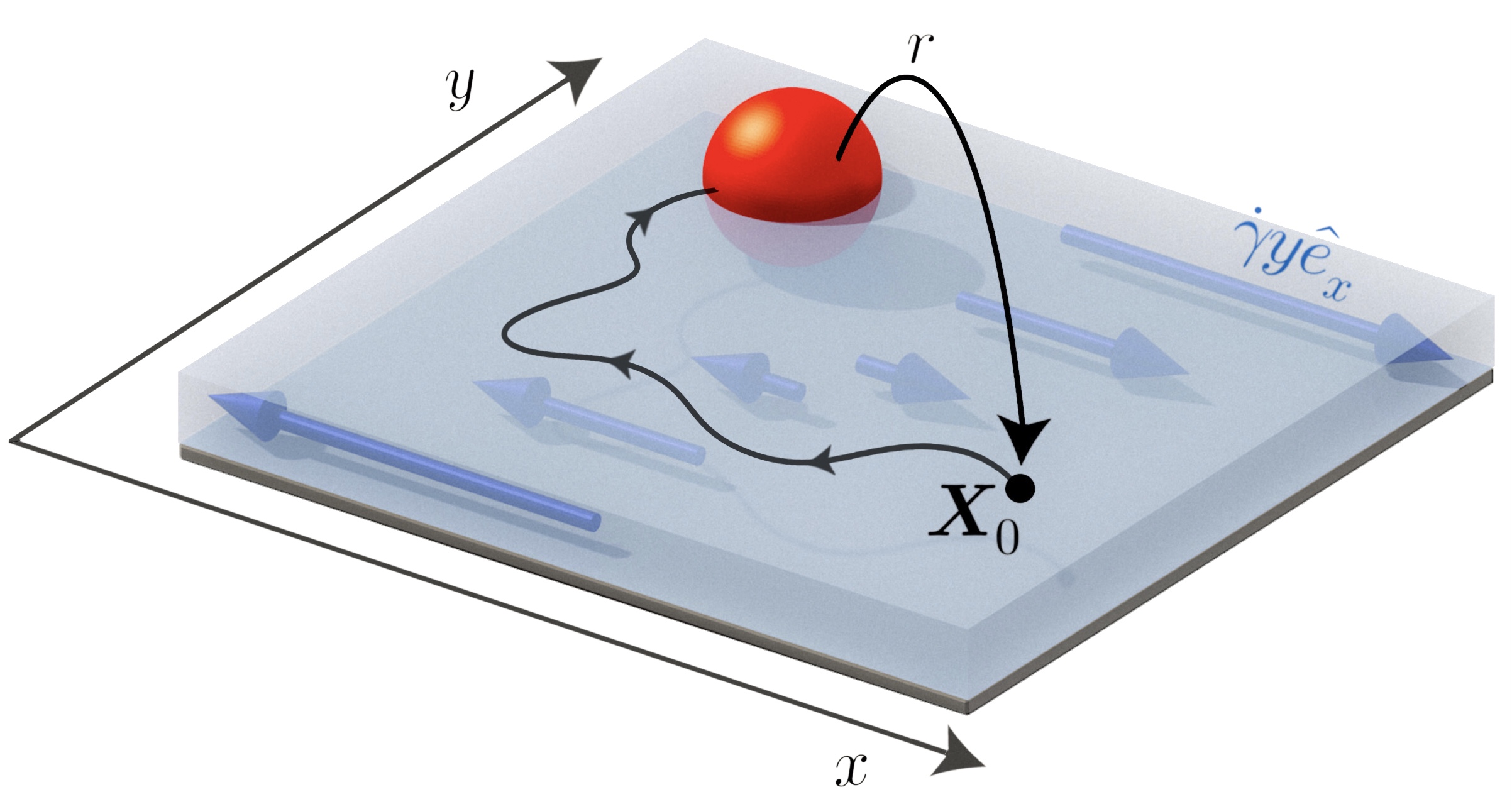}
    \caption{ 
Schematic of a two-dimensional Brownian particle in a linear shear flow in the $x$-direction $\dot \gamma y\hat{e}_x$ where $\dot \gamma$ is the shear rate. The particle undergoes stochastic resetting to its initial position $\mathbf{X}_0=(x_0, y_0)$ at random times with a constant rate $r$, indicated by the arched arrow.}
    \label{fig:schema01}
\end{figure}

\section{Model}\label{sec:back}
We consider a two-dimensional Brownian particle subject to a combination of shear flow and stochastic resetting. The particle diffuses with a constant diffusion coefficient $D$ while simultaneously being advected in a linear shear flow characterized by a shear rate $\dot{\gamma}$, where the velocity in the $x$-direction is proportional to the $y$-coordinate. At random intervals, the particle is reset to its initial position $(x_0, y_0)$ with a constant resetting rate $r$. The waiting time between two consecutive resetting events is a random variable with a Poisson distribution: in
a small time interval $\Delta t$ the particle is either reset to its initial
position with probability $r\Delta t$ or continues to diffuse with
probability $1-r\Delta t$. An illustration of the system is shown in Fig.~\ref{fig:schema01}.

The probability density for finding the particle at position $(x, y)$ at time $t$, given that it started at and reset to $(x_0, y_0)$, $p(x, y, t|x_0, y_0)\equiv p(x, y, t)$ is governed by the following Fokker-Planck equation (FPE) \cite{evans2020review}


\begin{align}
    \partial_t p(x,y,t) & = D \nabla^2 p (x,y,t)- \dot \gamma y\partial_x p(x,y,t) \nonumber \\
    & - rp (x, y, t) + r\delta(x-x_0)\delta(y-y_0),
    \label{eq:full_fpe}
\end{align}
where $\partial_i$ stands for derivative with respect to $i\in\{t, x\}$.
The first term on the right hand side represents pure diffusion and the second term corresponds to the advection in linear shear flow. The resetting mechanism is modeled by introducing a loss term, proportional to the resetting rate $r$, that removes probability density from all positions and redistributes it at the resetting position $(x_0, y_0)$, which are represented by the third and forth terms, respectively. 

The time-dependent FPE without resetting (i.e., ignoring the third and the forth terms) can be solved (see Appendix~\ref{appendix_A}) and the solution reads~\cite{novikov1958concerning,elrick1962source}

\begin{equation}\label{eq:pdf_t}
    p(x,y,t) = \frac{\sqrt{3}}{2 \pi D t \sqrt{ (12 + (\dot \gamma t)^2)}} e^{-\phi(x,y,t)},
\end{equation}
where we defined
\begin{align}\label{eq:phi}
   & \phi(x,y,t) = \frac{\left(y-y_0\right){}^2 \left(\dot\gamma^2 t^2+3\right)+3 (x-\dot\gamma t y)^2}{D t (12 + (\dot\gamma t)^2)} \\
   &+ \frac{3 \left(\dot\gamma t \left(y\!-\!y_0\right)\!-\!2 x_0\right) (x\!-\!\dot\gamma t y)\!+\!3 \dot\gamma t x_0
   \left(y_0\!-\! y\right)+3 x_0^2}{D t (12 + (\dot\gamma t)^2)}.\nonumber
\end{align}

\section{Non-equilibrium steady state}\label{sec:ness}
\begin{figure}[t]
	\centering
	\includegraphics[width=8.6cm]{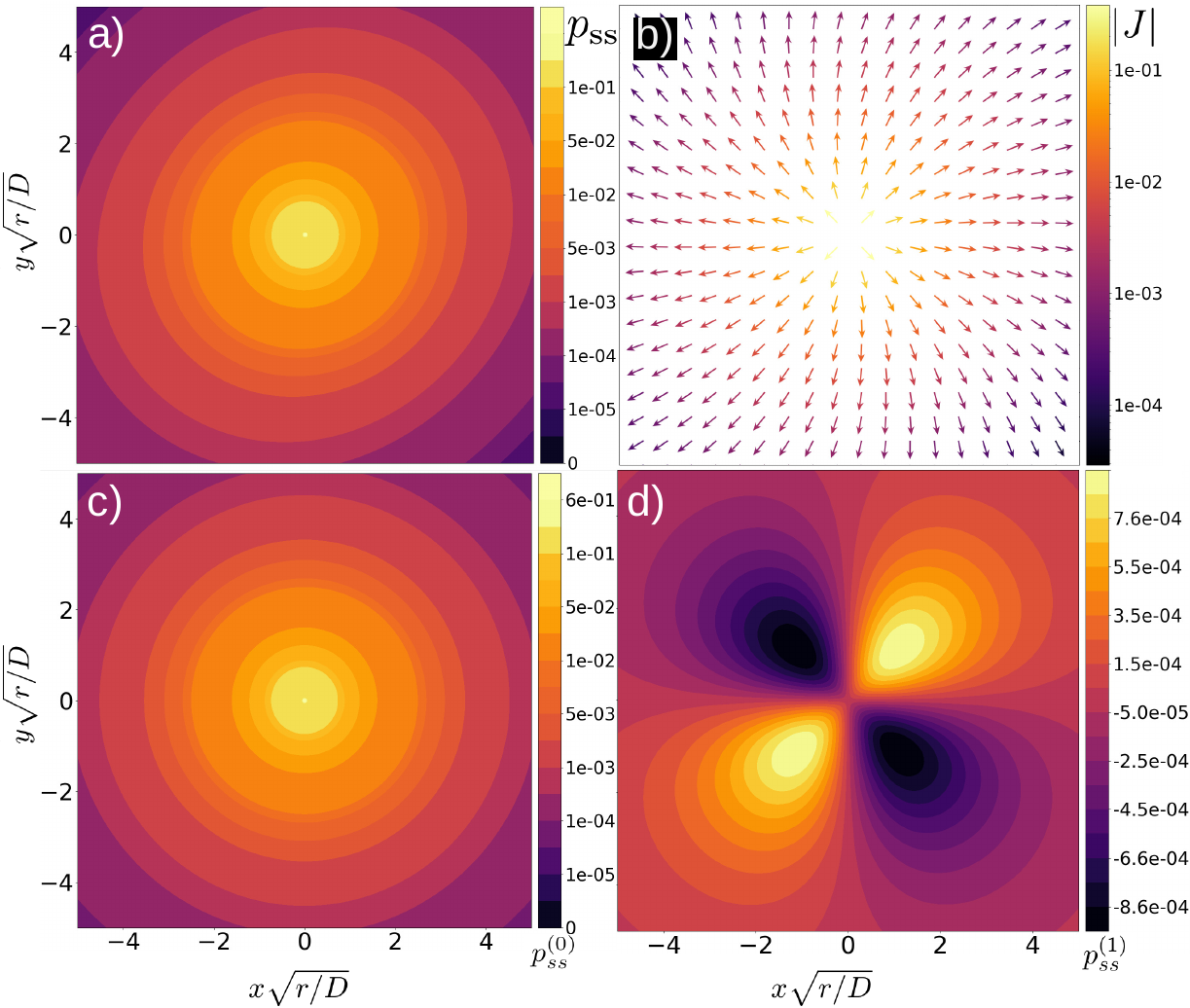}
	\caption{The steady-state probability density of the particle’s position and the corresponding probability fluxes in the system obtained from Eq.~\eqref{eq:pdf_ss} and Eq.~\eqref{eq:fluxes} for $\dot\gamma/r = 0.1$ are shown in (a) and (b), respectively. The direction of the fluxes is shown by the arrows; the magnitude is color-coded. (c) shows the diffusive core (without shear) and (d) represents the first order correction which induces anisotropy in the spatial structure of the distribution.  The particle starts at and resets to $(0.0, 0.0)$.}
	\label{fig:pdf_perturbative}
\end{figure}

The  probability density under stochastic resetting can be computed by using the renewal approach.  The probability density in the presence of resetting, denoted by $p_r(\bm{X},t|\bm{X}_0)$, can be obtained from the renewal equation  \cite{evans2020review} 
\begin{align}\label{eq:renewal_full}
	p_r(\bm{X}&,t|\bm{X}_0) = e^{- r t}p(\bm{X},t|\bm{X}_0) \\
	&+ r\int_0^t d\tau e^{-r\tau} \int d\bm{Y} p_r(\bm{Y},t-\tau|\bm{X}_0) p(\bm{X},\tau|\bm{X}_R),\nonumber
\end{align}
where $\bm{X} = (x,y)$ and $\bm{X}_R$ is the resetting location. The first term corresponds to trajectories where no resetting took place. The second term takes into account trajectories (with resetting) up to the time of the last resetting event before time $t$, i.e. at time $t-\tau$, when the particle is at position $\bm{Y}$. After the last reset, the particle propagates from the resetting location to $\bm{X}$ in the remaining time $\tau$. 

The resetting process interrupts the particle's natural diffusive trajectory, leading to a non-equilibrium steady state. In the steady state, the renewal equation simplifies to

\begin{align}\label{eq:renewal_ss}
	p_\text{ss}(\bm{X}|\bm{X}_0) &= r\int_0^\infty d\tau e^{-r\tau}  p(\bm{X},\tau|\bm{X}_R),
\end{align}
where \(p(\bm{X},\tau|\bm{X}_R)\) is given in Eq.\eqref{eq:pdf_t} and Eq.\eqref{eq:phi}. Obtaining an exact solution to the above equation is challenging, so we use a perturbative approach to solve it (see Appendix~\ref{appendix_B} for details). The solution reads

\begin{align} \label{eq:pdf_ss}
	p_\text{ss}(x, y) \approx \left( \frac{r}{2 \pi D} - \frac{\dot{\gamma} r \left[ x(y - y_0) - 2 x y \right]}{8 \pi D^2} \right) \nonumber \\
	\times K_0 \left( \alpha \sqrt{(y - y_0)^2 + x^2} \right),
\end{align}
where $\alpha=\sqrt{r/D}$. When no shear flow is included, i.e. $\dot \gamma = 0$, we recover the results for two-dimensional Brownian motion under resetting \cite{evans2014diffusion}.  

Using the steady-state probability density in Eq.~\eqref{eq:pdf_ss}, we can calculate the probability fluxes in the system as

\begin{equation}
	\label{eq:fluxes}
	\mathbf{J}(x, y) = -D\nabla p_{ss}(x, y) + \mathbf{v}(x, y)p_{ss}(x, y),
\end{equation}
where $\mathbf{v}(x, y)=(\dot \gamma y, 0)$ is the drift velocity due to the shear flow, which acts in the $x$-direction and depends linearly on $y$. The expression for the fluxes is given in Appendix~\ref{appendix_B}. 

\begin{figure}[t]
	\centering
	\includegraphics[width=8.6cm]{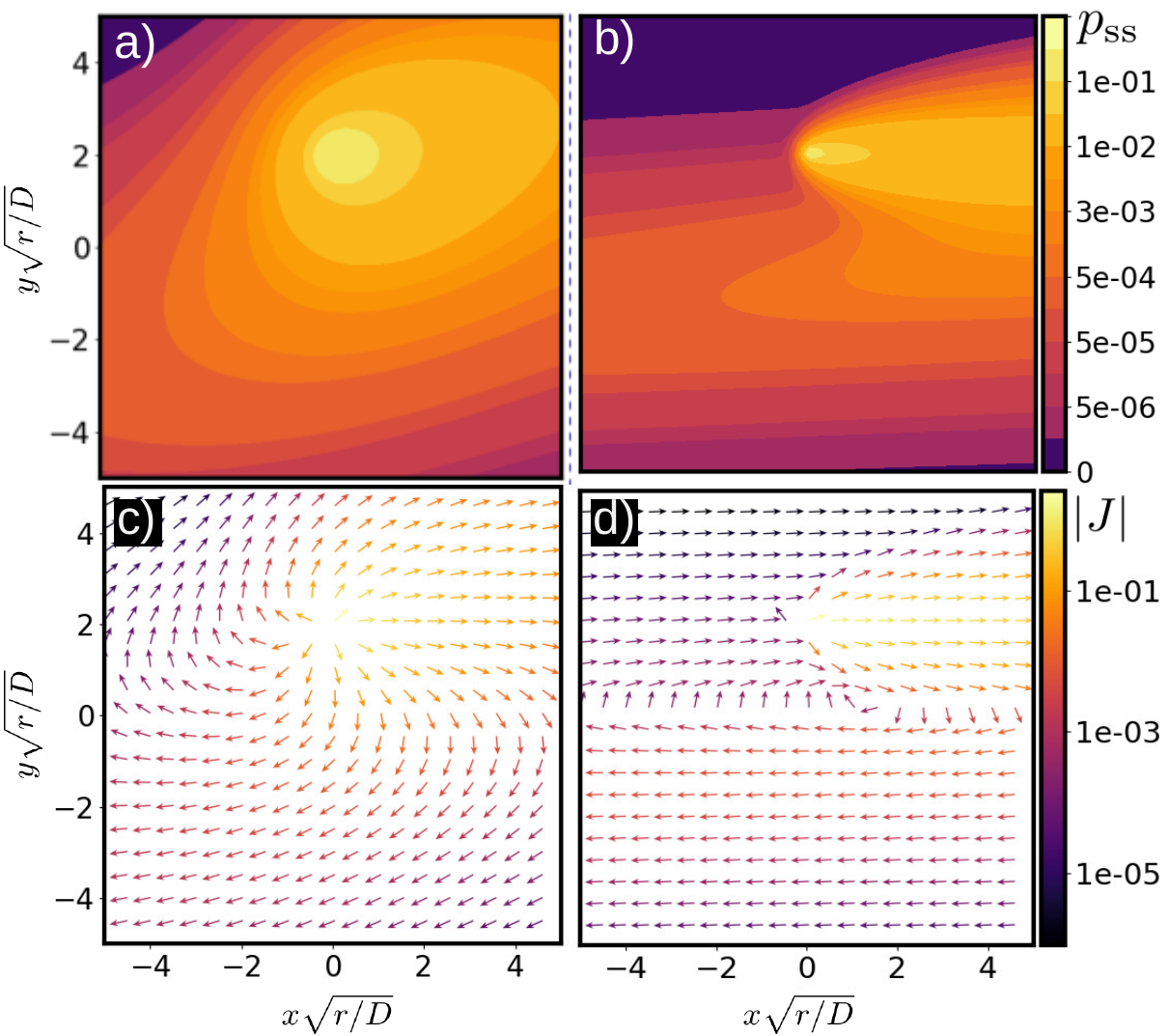}
	\caption{ The steady-state probability density of the particle’s position and the corresponding probability fluxes: (a) and (c) in a system with $\dot\gamma/r = 1.0$, (b) and (d) in a system with $\dot\gamma/r = 10.0$. The particle starts at and resets to $(0.0, 2.0)$. For such large shear rates the distribution becomes further stretched as going away from vertical initial position. The results are obtained by numerically solving Eq.~\eqref{eq:renewal_ss} and Eq.~\eqref{eq:pdf_t}. The direction of the fluxes is shown by the arrows; the magnitude is color-coded.}
	\label{fig:pdf_numerics}
	\end{figure}
Figure~\ref{fig:pdf_perturbative} (a) and (b), respectively, show the steady-state probability distribution of the particle's position from Eq.~\eqref{eq:pdf_ss} and the corresponding fluxes in the system from Eq.~\eqref{eq:fluxes} for a small value of the shear rate. The system is characterized by a diffusive core, shown in (c), with a correction term that reflects the influence of the shear flow, which is represented in (d). This correction term introduces anisotropy into the spatial structure of the distribution, leading to different behaviors in the two spatial directions. Figure~\ref{fig:pdf_numerics} shows the results for larger values of the shear rate from the numerical solution of Eq.~\eqref{eq:renewal_ss}, Eq.~\eqref{eq:pdf_t}, and Eq.~\eqref{eq:phi}.
  
\subsection{Moments}

Further insights can be obtained by considering the moments and cumulants. From the renewal equation, any observable $\mathcal{O}(x,y)$ can be studied under resetting through
\begin{align}\label{eq:renewal_obs}
	\langle \mathcal{O}(t)|\mathbf{X}_0\rangle_r = e^{- r t}\langle \mathcal{O}(t)|\mathbf{X}_0\rangle + r\int_0^t d\tau e^{-r\tau}  \langle \mathcal{O}(\tau)|\mathbf{X}_R\rangle.
\end{align}
Here the expectation values without subscripts are calculated using the time-dependent solution in Eq. \eqref{eq:pdf_t} and Eq.\eqref{eq:phi}. Below we consider several moments, cumulants and cross correlations to better understand the competing effects of shear and resetting. Details regarding the moments of the system without resetting are given in the Appendix~\ref{appendix_A}, which we here use in conjunction with the above renewal equation, Eq. (\ref{eq:renewal_obs}).

The centered means are simply given as
\begin{align}\label{eq:means}
	\langle x-x_0\rangle_\text{ss} &=  \frac{\dot \gamma y_0}{r},\\ 
	\langle y-y_0\rangle_\text{ss} &= 0 . \label{eq:means_y}
\end{align}
While the process is centered in the $y$-direction, there is a bias in the $x$-direction governed by the shear rate and the initial position $y_0$. The variances read
\begin{align}\label{eq:variances}
	\langle [x(t) - \langle x(t) \rangle ]^2 \rangle_\text{ss}  &= \frac{2 D (r^2 + 2 \dot \gamma^2) }{r^3}  + \frac{\dot\gamma^2}{r^2}y_0^2, \\ 
	\langle [y(t) - \langle y(t) \rangle ]^2 \rangle_\text{ss}  &= \frac{2 D}{r}.
\end{align}
\begin{figure}[t]
	\centering
	\includegraphics[width=8.6cm]{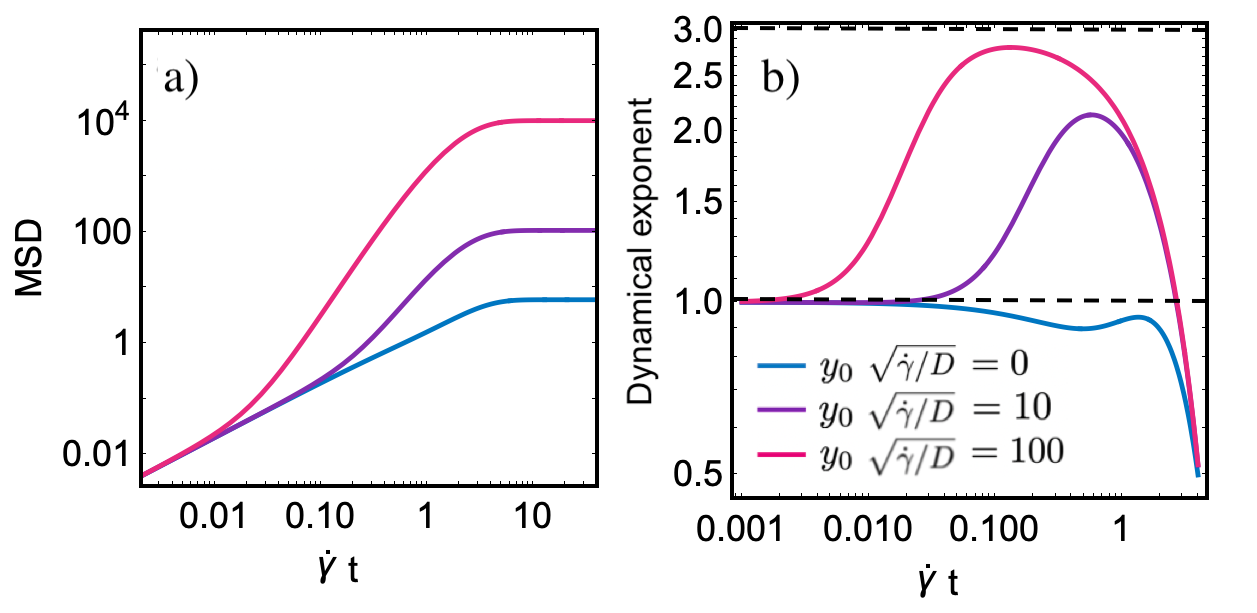}
	\caption{ (a) The mean square displacement in units of $\sqrt{D/\dot\gamma}$ in the $x$-direction and (b) the corresponding dynamical exponent with respect to (dimensionless) time for different values of the (dimensionless) initial position $y_0\sqrt{\dot\gamma/D}$. While the shear rate gives rise to a monotonic increase of the variance, the resets confine the steady state and makes the variance smaller. When the particle starts its motion away from the origin, the dynamics start from normal diffusion and cross over to super-diffusion due to the shear flow, before resetting finally brings the system to a steady state. The crossover ceases to exist in the absence of the shear flow. }
	\label{fig:msd_t}
\end{figure}
Since the motion is purely diffusive in the $y$-direction, the variance $ \langle [y(t) - \langle y(t) \rangle ]^2$ is identical to that of ordinary Brownian motion. In the $x$-direction, the steady-state variance is more complex and has a crossover
\begin{equation}
	\langle [x(t) - \langle x(t) \rangle ]^2 \rangle_\text{ss}   \propto
	\left\{
	\begin{array}{ll}
		r^{-3}  & \mbox{if } r  \ll \dot \gamma, \\
		r^{-1} & \mbox{if }  r  \gg \dot \gamma.
	\end{array}
	\right.
\end{equation}
Since the steady-state variance for Brownian particles under resets normally scale as $\sim r^{-1}$, we can interpret the above new scaling $\sim r^{-3}$ as the regime in which the shear flow plays a dominant effect. The crossover resetting rate $r_c$ is given by matching the small- and large-$r$ behaviors, resulting in $r_c = \sqrt{2} \dot \gamma$. Hence, there is no crossover and only one scaling regime in the absence of shear ($\dot \gamma = 0 \Rightarrow r_c = 0$), while for infinitely strong shear only the $\sim r^{-3}$ regime can be observed. We also note that while the shear rate gives rise to a monotonic increase of the variance, the resets confine the steady state and makes the variance smaller. Figure.\ref{fig:msd_t} (a) shows the mean square displacement (MSD) as a function of time, with panel (b) showing the dynamical exponent $\zeta(t) = \frac{\partial}{\partial \log t} \log \langle [x(t) - \langle x(t) \rangle ]^2 \rangle$ which governs the typical temporal scaling $\langle [x(t) - \langle x(t) \rangle ]^2 \rangle\sim t^{\zeta}$. Clearly, multiple dynamical crossovers exist depending on the value of $y_0$. At early times, the motion is diffusive. For $y_0>0$ the dynamics cross over to super-diffusive due to the shear flow, before resetting finally brings the system to a steady state. When $y_0=0$, the steady state is approached in a purely sub-diffusive manner.

The steady states also gain non-zero cross correlations due to the shear flow. This can be measured by the first-order cross-correlation function
\begin{equation}\label{eq:cross_correlations}
	\langle (x- \langle x\rangle)(y-y_0)\rangle_\text{ss} = \frac{2 D \dot \gamma}{r^2},
\end{equation}
which clearly vanishes when $\dot \gamma = 0$. Note that $\langle y\rangle = y_0$ from Eq.\eqref{eq:means_y}.

\begin{figure}[t]
	\centering
	\includegraphics[width=8.6cm]{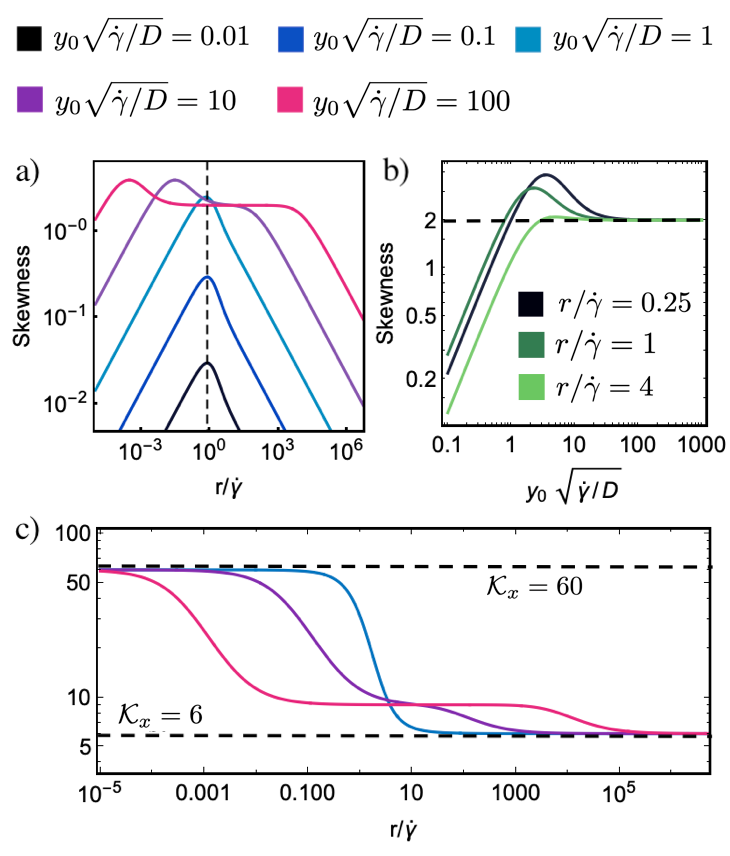}
	\caption{ 
	The steady-state skewness and kurtosis in the $x$-direction, with lengthscales set by $\sqrt{D/\dot\gamma}$, and timescales by $\dot\gamma^{-1}$. (a) shows the skewness with respect to the dimensionless resetting rate $r/\dot\gamma$. The skewness performs a non-monotonic behavior as one passes from the shear-dominated ($r/\dot\gamma \ll 1$) to the resetting-dominated ($r/\dot\gamma \gg 1$) regime. When the problem is symmetric around $x=0$, the skewness is zero. The dashed line shows the optimal resetting rate, $r^\star/\dot\gamma \approx 0.764$ that maximizes the skewness for small $y_0$. (b) indicates that, independent of the dimensionless parameter $r/\dot\gamma$, the skewness saturates to $2$ when the particle starts its motion at $y_0\gg 1$. As shown in (c), the kurtosis is $60$ where the shear is dominated the resetting and decreases monotonically with increasing $r/\dot\gamma$ saturating to $\mathcal{K}_x=6$ where resetting dominates shear.   }
	\label{fig:skew_kurt}
\end{figure}

The skewness $\mathcal{S}_x$ can in the steady state be calculated as
\begin{align}
	\mathcal{S}_x(r) &= \lim_{t\to \infty}\frac{\langle (x-\langle x \rangle)^3\rangle}{\langle (x-\langle x \rangle)^2\rangle^{3/2}} \\
	&= y_0 \frac{\sqrt{r}  \left[6 \dot{\gamma } D \left(6 \dot{\gamma }^2+r^2\right)+2 \dot{\gamma }^3 r
		y_0^2\right]}{\left(2 D \left(2 \dot{\gamma }^2+r^2\right)+\dot{\gamma }^2 r y_0^2\right)^{3/2}}. \label{eq:skew_x}
\end{align}
While the skewness has a complex behavior as the resetting rate is varied, a general feature is that it vanishes for $y_0=0$, when the problem is symmetric around $x=0$. Furthermore, to leading order is small resetting rates $r$, the skewness grows as $\mathcal{S}_x(r)\sim r^{1/2}$. However, at late times the skewness decays as $\mathcal{S}_x(r)\sim r^{-1/2}$, clearly showing a non-monotonic behavior as one passes from the shear-dominated to the resetting-dominated regime. Intuitively, when resetting dominates, the steady state approached the shear-free system which is symmetric. When shear dominates, the steady state is stretched out and becomes flatter, also decreasing the skew. In Fig.~\ref{fig:skew_kurt} (a), we show the non-monotonicity of the skewness for different values of $y_0$ where we use Eq.~\eqref{eq:skew_x} for the plots. 
	
Identifying the value of the resetting rate that maximizes the skew is arduous for general parameter regimes. However, to leading order in $y_0$, as the system is perturbed away from its symmetric conditions, one can identify the optimal resetting rate 
	\begin{equation}
		\frac{r_*}{\dot \gamma} = \sqrt{(4\sqrt{7}-10)}\approx 0.764...
	\end{equation}
This dimensionless value characterizes the balance between the strengths of resetting and shear, which is shown in Fig.~\ref{fig:skew_kurt} (a) by the dashed line.

As a function of the initial vertical displacement $y_0$, however, the skewness also shows a non-trivial behavior. While the steady state has zero skew for $y_0=0$ and initially increases linearly as $y_0$ is increased, it saturates at the value $\mathcal{S}_x(y_0\to \infty) = 2$. This is independent from the value of $r/\dot\gamma$ as shown in Fig.~\ref{fig:skew_kurt} (b). An optimal value of $y_0$ can be found, taking the form 
	\begin{equation}
		y_0^* = \frac{1}{2}\sqrt{\frac{D}{r} \left(12 + \left(\frac{r}{\dot\gamma}\right)^4+8 \left(\frac{r}{\dot\gamma}\right)^2\right)}.
	\end{equation}

The kurtosis in the $x$-direction, which measures the tailedness or non-Gaussianity of the marginal distribution, can be calculated to be 
\begin{align}
	\mathcal{K}_x &\equiv \lim_{t\to \infty }\frac{ \langle [x(t) - \langle x(t) \rangle ]^4 \rangle }{ \langle [x(t) - \langle x(t) \rangle ]^2 \rangle^2}\\
	& = { \frac{3 \left[8 D^2 \left(40 \dot{\gamma }^4+r^4+8 \dot{\gamma }^2 r^2\right)\right]}{\left[2 D \left(2
			\dot{\gamma }^2+r^2\right)+\dot{\gamma }^2 r y_0^2\right]^2}} \nonumber \\
	& + { \frac{3 \left[4 \dot{\gamma }^2 D r
			y_0^2 \left(26 \dot{\gamma }^2+3 r^2\right)+3 \dot{\gamma }^4 r^2 y_0^4\right]}{\left[2 D \left(2
			\dot{\gamma }^2+r^2\right)+\dot{\gamma }^2 r y_0^2\right]^2}}.
		\label{eq:kurtosis}
\end{align}

In Fig.~\ref{fig:skew_kurt} (c), using the above equation we plot the kurtosis. It clearly shows that  $\mathcal{K}_x=60$ where shear dominates resetting and decreases monotonically with increasing $r/\dot\gamma$ saturating to $\mathcal{K}_x=6$ where resetting dominates shear. This is the result for the Laplace distribution known to be the steady state for one-dimensional diffusion with resets.

\subsection{Cost of maintaining the steady state}

In recent years, the cost needed to maintain the non-equilibrium steady state resulting from resetting (often a thermodynamic cost such as entropy or work) has been studied intensively, both in theory and experiment \cite{fuchs2016stochastic,olsen_thermodynamic,olsen2024thermodynamic,mori2023entropy,Deepak2022_work,gupta2020work,Busiello2020uni,pal2023thermodynamic,sunil2023cost,Roberts_2024,goerlich2024resetting,singh2024cost,sunil2024minimizing,tal2024smart}. This not only gives insights into how far from equilibrium the system is, but also provides a measure of the energetic cost associated with performing random recurrent resets using, e.g., optimal tweezer setups. Here we consider, in the simplest scenario, the mean thermodynamic work associated with resetting in the presence of shear flow. 

\begin{figure}[t]
	\centering
	\includegraphics[width=\columnwidth]{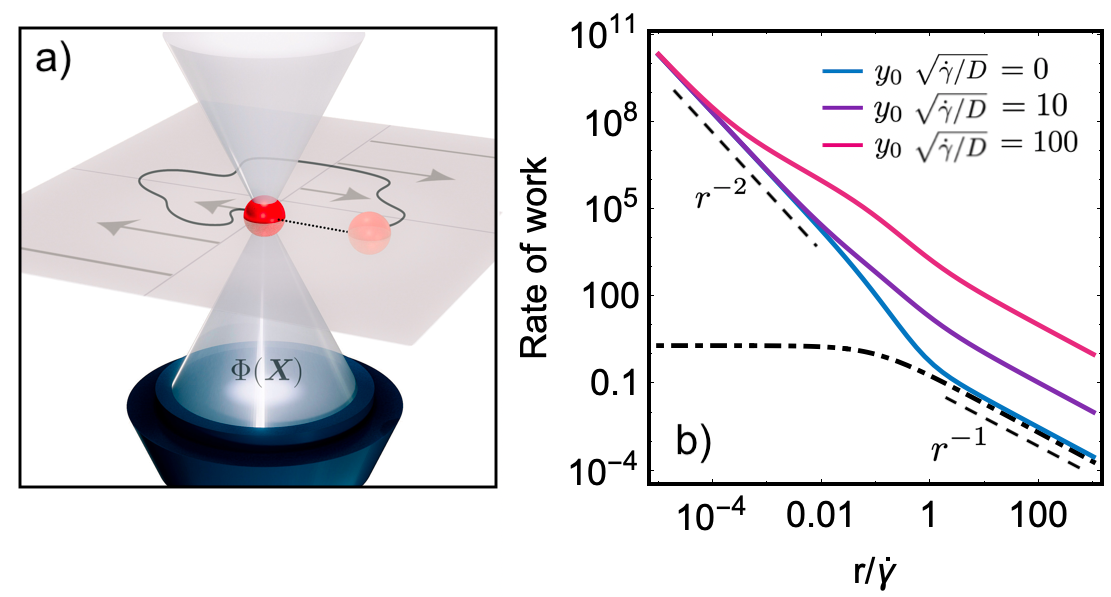}
	\caption{ (a) Resetting can be implemented by a trap generated, for example, by an optical tweezer. (b) Each reset comes at a thermodynamic cost, which can be measured by the work. Solid lines show the rate of work for various values of $y_0$, while the (curved) dashed line shows the rate of work in the absence of shear flow. Even though resetting events are infrequent at small resetting rates, each one results in a substantial energetic cost. The results are from Eq.\eqref{eq:work} to Eq:\eqref{eq:mu_SD}. Dimensionless units are used, where lengthscales are set by $\sqrt{D/\dot\gamma}$, and timescales by $\dot\gamma^{-1}$. Parameters are set to $\lambda/\dot\gamma = 1$ and $\dot \gamma \tau_R = 10$.}
	\label{fig:work1}
\end{figure}

The thermodynamic cost of resetting depends on the specific implementation of the resetting mechanism, which hitherto we have left unspecified. In the simplest case, a confining trap $\Phi(\bm{X})$, with a minimum at the resetting location, is switched on at the time of reset, and deactivated after the particle has relaxed near the trap minimum (see Fig. \ref{fig:work1}). This was explored in  Ref. \cite{olsen_thermodynamic}, which we follow here. The energetic cost associated with activating the trap in the steady-state regime is simply $\langle \Phi(\bm{X})|\bm{Y}\rangle_\text{ss}$. Since the particle relaxes to the steady state in the trap at every resetting event, the expectation value is conditioned on $\bm{Y}$, which is a random variable distributed with the steady state $P_\text{trap}(\bm{Y})$ associated with the combined effect of a potential and shear flow. Once the potential is deactivated, the energy cost associated with a full resetting cycle is $\langle \Phi(\bm{X})|\bm{Y}\rangle_\text{ss}-\Phi(\bm{Y}))$.

If the time during which the potential is active is denoted by $\tau_R$ (which must be larger than the relaxation time of the potential), then the relation between observation time $t$ and mean number of resets $\overline{n}(t)$ is $t = (r^{-1}+\tau_R) \overline{n}(t)$. Hence, a total mean work $\langle W (t)|\bm{Y}\rangle =\left[\langle \Phi(\bm{X})|\bm{Y}\rangle_\text{ss}-\Phi(\bm{Y}))\right] \overline{n}(t)$ must be paid. The steady state rate of mean work 
\begin{align}
    \mu_\text{W}(r) &= \lim_{t\to \infty} \frac{\langle W \rangle}{t} \nonumber \\
    &= \frac{\int d\bm{Y} P_\text{trap}(\bm{Y}) \left[\langle \Phi(\bm{X})|\bm{Y}\rangle_\text{st} - \Phi(\bm{Y})\right]}{r^{-1}+\tau_R}, \label{eq:rateofwork}
\end{align}
is therefore a reasonable measure of the thermodynamic cost needed to maintain the steady state \cite{olsen_thermodynamic,olsen2024thermodynamic}. 

As an example, we assume that the resets are mediated by a harmonic trap $\Phi(\bm{X}) = \frac{1}{2}\lambda (\bm{X}-\bm{X}_R)^2$, where $\lambda$ is the stiffness of the potential and the resetting location is $\bm{X}_R = (0,y_R)$. Moments associated with $P_\text{trap}(\bm{Y})$ are needed for the calculation of the rate in Eq. (\ref{eq:rateofwork}), which we report in Appendix \ref{appendix_C}. Combining this with the previous results of section \ref{sec:ness}, we find a rate of work that has three contributions, namely
\begin{equation} \label{eq:work}
   \mu_\text{W}(r)  = \mu_\text{D}(r)+\mu_\text{S}(r)+\mu_\text{SD}(r),
\end{equation}
where
\begin{align} \label{eq:mu_D}
    \mu_\text{D}(r) &= \frac{2 D \lambda}{1+r \tau_R}, \\ 
    \mu_\text{S}(r) &= \frac{\dot\gamma^2 y_R^2 (r+\lambda)}{ r (1+r\tau_R)}, \label{eq:mu_S} \\ 
    \mu_\text{SD}(r) &= \frac{ D \dot\gamma (r+2\lambda)}{ r^2(1+r\tau_R)}. \label{eq:mu_SD}
\end{align}
We notice that $\mu_\text{D}(r)$ is independent of the shear rate, and therefore originates purely from diffusion.  When $\dot\gamma = 0$, we recover the expected results $\mu_\text{W}(r)=\mu_\text{D}(r)$ as reported in Ref. \cite{olsen_thermodynamic} up to a numerical factor owing to the fact that we work in two dimensions rather than one. The contribution $\mu_\text{S}(r)$ is independent of the diffusivity, and comes from the vertical resetting location in the shear flow. Lastly, the third term $\mu_\text{SD}$ has mixed origins. 

At small resetting rates, we have the leading order behavior
\begin{equation}
    \mu_\text{W}(r) \approx \frac{2 D\dot\gamma^2 \lambda}{r^2}.
\end{equation}
This indicates that while a resetting event is very rare in this limit, once it occurs there will be a large energetic cost. This is a regime not observed in the absence of shear flow. Indeed, it was pointed out in Ref. \cite{olsen_thermodynamic} that in the absence of any background flow, the mean rate of work for resets that are carried out with a harmonic resetting trap is independent of $r$ at small $r$-values due to competing effects; 1) rare resetting events cause the work to decrease, while 2) eventual resets will come at a high cost since the particles has had time to diffuse far away. In the present case, these two effects are no longer in balance, since the flow transports the particle further than what it would reach by pure diffusion alone. This causes the eventual cost of a reset to be much higher, hence the $\sim r^{-2}$ scaling observed at small $r$, which is shown in Fig.~\ref{fig:work1} (b).

As discussed in the preceding sections, the system experiences a competition between shear, which promotes skewness and cross-correlations, and resetting, which acts to restore (parity) symmetry. This competition can have interesting consequences for the rate of work needed to produce a steady state with certain asymmetry properties. In the following, we assume that resets can be performed with a sharp trap that well approximates the instantaneous resetting used in preceding sections. 

As most observables show crossovers between different scaling behaviors as a function of the resetting rate, it is convenient to consider the resetting-dominated and shear-dominated regimes separately. At small rates, one we can show by combining the above results that
\begin{equation}
	\mathcal{S}_x^4(r) \mu_W(r) \approx \frac{\dot \gamma^2 \lambda y_R^4}{D} \sim \text{const.}
\end{equation}
which is independent of rate $r$. This relation is a trade-off relation that states that if we want to tune the rate of resets to increase skewness, the rate of work goes down (and vice versa). For example, doubling the skewness leads to a rate of work reduced by a factor of 16. Analogously, in the resetting-dominated regime at high rates, we have 
\begin{equation}
	\frac{\mu_W(r)}{\mathcal{S}_x^2(r)} \approx \frac{2 D [\dot \gamma^2(y_R^2 + D/\lambda) + 2 D \lambda]}{9 \dot \gamma^2 y_R^2 \tau_R} \sim \text{const.}
\end{equation}
In contrast to the shear-dominated regime, here we see that tuning the rate to increase the skewness requires an increase also in the rate of work.    

\begin{figure}[t]
    \centering
    \includegraphics[width=8cm]{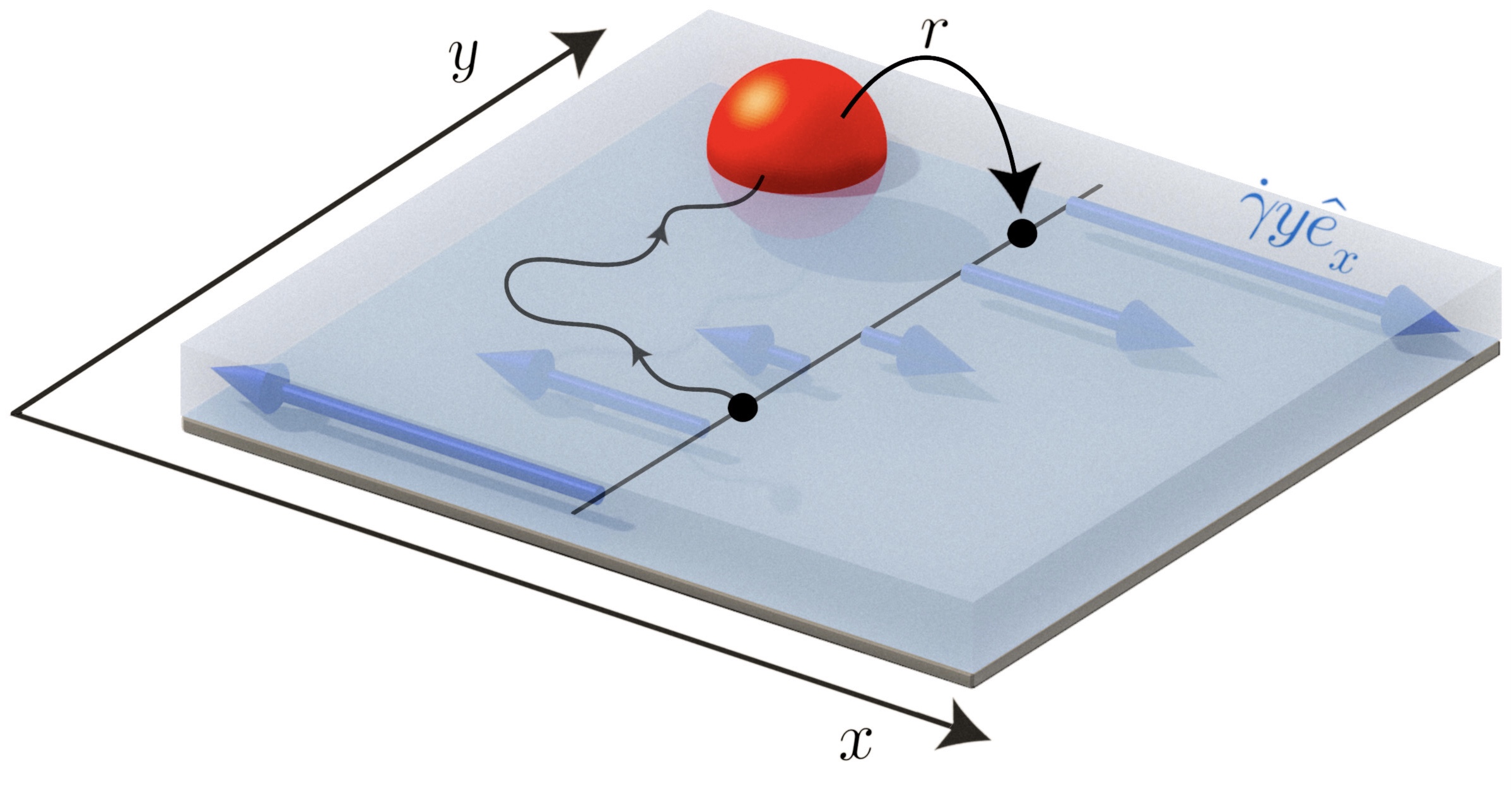}
    \caption{Schematic of a two-dimensional Brownian particle in a linear shear flow in the $x$-direction $\dot \gamma y\hat{e}_x$ where $\dot \gamma$ is the shear rate. The $x$-coordinate of the particle is reset, which is indicated by the arched arrow, while the $y$-coordinate is diffusing freely.  }
    \label{fig:xreset}
\end{figure}
\section{Resetting parallel to the flow}
Resetting famously gives rise to steady states by confining a system's trajectories. For a pure diffusion process in the plane, a steady state in the $x$-direction is obtained even if the $y$-coordinate is left to diffuse freely and only $x$ is reset. In the present case, however, the effect of advection increases as $y$ is allowed to grow. A priori, it is not clear whether the $x$-dynamics will reach a steady state if only the particles $x$ component is reset (see Fig. \ref{fig:xreset}).

In this case, the last renewal equation takes the form
\begin{align}
    p_r(&\bm{X},t|\bm{X}_0) = e^{- r t}p(\bm{X},t|\bm{X}_0) \nonumber\\
    & + r \int_0^t d\tau e^{- r \tau}\int d \bm{X}' p_r(\bm{X}',t-\tau|\bm{X}_0) p (\bm{X},\tau|0,y').
\end{align}
Since we are interested in the dynamics of the $x$ coordinate, we can integrate out $y$. This gives
\begin{align}
    \rho_r(&x,t|\bm{X}_0) = e^{- r t}\rho(x,t|\bm{X}_0) \nonumber\\
    & + r \int_0^t d\tau e^{- r \tau}\int d y' \wp_r(y',t-\tau|\bm{X}_0) \rho (x,\tau|0,y'),
\end{align}
where we defined the marginal densities
\begin{align}
    \rho_r(x,t|x_0,y_0) &\equiv \int dy p_r(x,y,t|x_0,y_0),\\
    \wp_r(y,t|x_0,y_0) &\equiv \int dx p_r(x,y,t|x_0,y_0).
\end{align}
Similar definitions hold without resetting. However, since we only reset $x$, the propagator $\wp_r(y,t|x_0,y_0)$ is unaffected by resetting. Furthermore, it does not depend on the initial $x$ coordinate. Hence, 

\begin{equation}
    \wp_r(y,t|x_0,y_0) = \wp(y,t|y_0),
\end{equation}
which is nothing but the standard Gaussian solution for a diffusion process with a point-source initialization at $y_0$. The propagator can be expressed as
\begin{align}
    \rho_r(x,t&|\bm{X}_0) = e^{- r t}\rho(x,t|\bm{X}_0) \nonumber\\
    & + r \int_0^t d\tau e^{- r \tau}\int d y' \wp(y',t-\tau|y_0) \rho (x,\tau|0,y').
\end{align}
The expectation value of any observable $\mathcal{O}(x,y)$ can as before be obtained from this renewal equation simply by multiplication by $\mathcal{O}(x,y)$ and integrating over $x$ and $y$.

Using the moments of the process without resetting given in Appendix \ref{appendix_A}, we find the horizontal variance
\begin{align} 
    \langle [x &-\langle  x \rangle]^2(t) \rangle = \nonumber \\
     &
   \frac{2 e^{-r t} \left(D r^2 \left(e^{r t}\!-\!1\right)\!+\!2 \dot{\gamma }^2 D \left[r t+e^{r t} (r t\!-\!2)\!+\!2\right]\right)}{r^3} \nonumber\\
   &+ \frac{2 e^{-r t} \left(\dot{\gamma }^2 r
   y_0^2 [\sinh (r t)-r t]\right)}{r^3}. \label{eq:msd_resetx}
\end{align}
As we show in Fig.~\ref{fig:msd_resetx}, several crossovers can be observed; the motion starts out by performing standard diffusion. This is followed by a superdiffusive regime, followed by a subdiffusive regime, before at late times diffusive behavior is once again recovered. The late-time diffusion coefficient can be found to be { 
\begin{equation}
    D_\text{eff} \equiv \lim_{t\to \infty} \frac{  \langle [x-\langle x \rangle]^2(t) \rangle}{2 t}= 2D \left( \frac{ \dot \gamma }{r} \right)^2.
\end{equation}
Surprisingly, even though the $x$-coordinate is reset at rate $r$, the system never reaches a steady state but rather spreads diffusively. A similar effect has been observed by  van den Broeck and coauthors when the particle's horizontal motion is confined by a harmonic trap while in a shear flow \cite{van1982harmonically}.
\begin{figure}[t]
	\centering
	\includegraphics[width=\columnwidth]{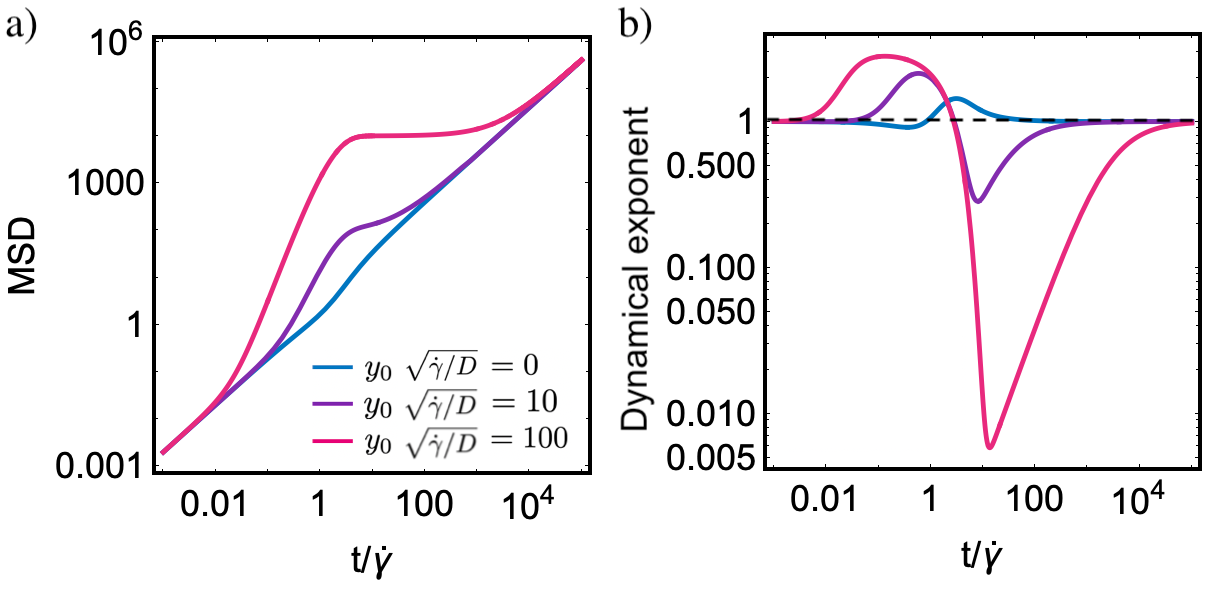}
	\caption{ 
		(a) The mean square displacement, for a shear-driven system where only $x$ is reset, in units of $\sqrt{D/\dot\gamma}$ in the $x$-direction and (b) the corresponding dynamical exponent with respect to (dimensionless) time for different values of the vertical initial position. The dynamics show complex behavior, with multiple crossovers between different regimes of diffusion, including super-diffusive and sub-diffusive phases before eventually returning to normal diffusive behavior. The results are obtained from Eq.~\eqref{eq:msd_resetx} }
	\label{fig:msd_resetx}
\end{figure}
While the dynamics is diffusive, the probability density itself is highly non-Gaussian. The kurtosis, at late times, can be calculated to be 
\(
	\lim_{t\to \infty}\mathcal{K}_x(t) = 18
\)
indicating a leptokurtic probability density of positions, where large fluctuations could occur. Intuitively, due to the lack of constraint in the vertical direction some particles will diffuse to high $y$-values, where the shear flow is very strong. These trajectories will contribute to large fluctuations in the positions beyond what is observed in normal Gaussian diffusion processes. Hence, the system is an example of Brownian yet non-Gaussian diffusion. While many such systems are found to have exponential tails (i.e., Laplacian spatial densities) corresponding to $\mathcal{K}_x =6$, we here observe a more extreme type of tail. It is also worth noting that this value is completely independent of the system parameters.  In particular, the late-time kurtosis is unaffected by the rate of shear or resetting, while the diffusion coefficient depends only on their ratio.

\section{Discussion}\label{sec:discussion}


In this work, we explored the non-equilibrium dynamics of a Brownian particle subject to both shear flow and stochastic resetting. One of the findings is the emergence of anisotropic steady-state distributions driven by the shear flow. The resetting mechanism, which typically leads to symmetric and confined steady states in simpler diffusive systems, is disrupted by the shear-induced asymmetry. This results in a skewed distribution that is particularly sensitive to the relative strengths of the shear flow and resetting rate. In the shear-dominated regime, we observed that the system develops substantial anisotropy, with the particle distribution stretching further along the direction of the shear. Conversely, in the resetting-dominated regime, the steady state regains its symmetry, as resetting overrides the effects of shear flow. This balance between shear and resetting is quantified through key statistical properties like skewness, which display non-monotonic behavior as the system transitions between these regimes.

Interestingly, when only the particle's $x$-coordinate is reset, we discovered that the system does not reach a steady state, despite the resetting mechanism. Instead, the particle's position continues to spread diffusively, suggesting that the advection due to shear prevents the confinement typically associated with resetting. This finding is particularly striking, as it highlights a case where resetting fails to establish a steady state, challenging the conventional understanding of resetting as a mechanism for system stabilization. The dynamics in this scenario show complex behavior, with multiple crossovers between different regimes of diffusion, including super-diffusive and sub-diffusive phases before eventually returning to normal diffusive behavior.

Furthermore, the energetic cost associated with maintaining the non-equilibrium steady state was examined. At low resetting rates, although resets are infrequent, they incur a disproportionately high energetic cost due to the particle's displacement under shear. This cost scaling with resetting rate differs markedly from the case without shear, where the cost remains relatively constant for low resetting rates. The sheared system, however, experiences a breakdown of this balance, leading to a substantial increase in the energetic cost when resets finally occur. This observation has practical implications for the design of resetting processes in systems where energetic efficiency is critical, such as in biological systems or optimization algorithms.

In summary, this work demonstrates that the inclusion of shear flow introduces significant complexity into the dynamics of diffusive systems with resetting. The observed behaviors underscore the importance of considering external forces when modeling resetting phenomena and provide new avenues for exploring how non-equilibrium steady states are formed and maintained. Future work could extend these findings by exploring resetting in more complex flow fields, such as turbulent or oscillatory flows, and examining how different forms of stochastic resetting (e.g., resetting to non-fixed positions or variable rates) influence the system's behavior. 

Finally, several recent experiments on stochastic resetting have taken place, most of which rely on optimal tweezer methods \cite{besga2020optimal,friedman2020exp,goerlich2023experimental}. Using similar methods, our predictions could be verified in experiments on sheared colloids.


\begin{acknowledgements}

The authors acknowledge funding by the Deutsche Forschungsgemeinschaft (DFG) within the project LO 418/29-1.
	
\end{acknowledgements}

\section*{AUTHOR DECLARATIONS}
\subsection*{Conflict of Interest}
The authors have no conflicts to disclose.
	
\section*{DATA AVAILABILITY}
The data that support the findings of this study are available
from the corresponding author upon reasonable request.
\appendix

\section{Time-dependent solution (without shear)} \label{appendix_A}
The probability for finding a Brownian particle under a linear shear flow in the $x$-direction at position $(x, y)$ at time $t$, given that it started at $(x_0, y_0)$, obeys the following Fokker-Planck equation

\begin{equation}
	\partial_t p(x,y,t) = - \dot \gamma y\partial_x p(x,y,t) + D \nabla^2 p (x,y,t)
\end{equation}
where we have suppressed the dependence on the initial positions $(x_0,y_0)$ for notational brevity. 

To solve it, we use coordinate transforms first proposed by Novikov and Elrick \cite{novikov1958concerning,elrick1962source}
\begin{align}
	u &= x- \dot\gamma y t,\\
	v &= y,\\
	q &= t,
\end{align}
which transforms the above Fokker-Planck equation into
\begin{equation}
	\partial_q  p = D  (1+\dot \gamma^2 q^2) \partial_u^2  p + D \partial_v^2 p - 2 D \dot \gamma q \partial_u \partial_v p.
\end{equation}
Performing a double Fourier transform, defined as
\begin{equation}
	\hat p(\xi,\eta,q) = \int du dv e^{- i \xi u - i \eta v}p(u,v,q),
\end{equation}
results in the Fourier-space solution
\begin{equation}
	\hat p(\xi,\eta,q) = e^{-\Lambda(\xi,\eta,q) - i \xi x_0 - i \eta y_0},
\end{equation}
where we have defined the function
\begin{equation}
	\Lambda(\xi,\eta,q) = D(\xi^2+\eta^2) q -  D\dot \gamma \xi \eta q^2+ \frac{1}{3} D\dot \gamma^2 \xi^2 q^3.
\end{equation}
Inverting the Fourier transform and transferring back to the original coordinates gives the propagator, which reads
\begin{equation}\label{eq:prop}
	p(x,y,t) = \frac{\sqrt{3}}{2 \pi D t \sqrt{12 + (\dot \gamma t)^2}} e^{-\phi(x,y,t)},
\end{equation}
where we defined
\begin{align} \label{eq:phi_app}
	& \phi(x,y,t) = \frac{\left(y-y_0\right){}^2 \left(\dot\gamma^2 t^2+3\right)+3 (x-\dot\gamma t y)^2}{D t (12 + (\dot\gamma t)^2)} \\
	&+ \frac{-3 \left(\dot\gamma t \left(y_0\!-\!y\right)\!+\!2 x_0\right) (x\!-\!\dot\gamma t y)\!+\!3 \dot\gamma t x_0
		\left(y_0\!-\!y\right)\!+\!3 x_0^2}{D t (12 + (\dot\gamma t)^2)}.\nonumber
\end{align}
 From this propagator, several observables can be calculated. The bare moments of lowest order are given by 
\begin{align}
    \langle x \rangle &=  x_0 + \dot\gamma y_0 t,\\
    \langle x y \rangle &=  x_0 y_0 + \dot\gamma t (y_0^2 + D t),\\
    \langle x^2 \rangle &=  (x_0 + \dot\gamma y_0 t)^2 + \frac{2}{3} D  [3 + (\dot \gamma t)^2] t,\\
    \langle x^3 \rangle &= (x_0 + \dot\gamma y_0 t)  \left[ 2 D t (3 + (\dot\gamma t)^2) + (x_0 + \dot\gamma y_0 t)^2 \right], \\
    \langle x^4 \rangle &= (x_0 + \dot\gamma y_0 t)^4 + (x_0 + \dot\gamma y_0 t)^2 4 D t (3 + (\dot\gamma t)^2),  \nonumber\\
    &+ \frac{4}{3} D^2 t^2 (3+ (\dot \gamma t)^2)^2.
\end{align}
More compactly, we can introduce the mean $\mu(t) = x_0 + y_0 \dot\gamma t$ and variance $\mathcal{V}(t) = \frac{2}{3} D  [3 + (\dot \gamma t)^2] t $, and write the centralized moments as
\begin{equation}
    \langle [x-\mu(t)]^n \rangle = \frac{2^{\frac{n-2}{2}}\Gamma \left(\frac{n+1}{2}\right)}{\sqrt{\pi}} \left[ 1 + (-1)^n \right]\mathcal{V}(t)^n,
\end{equation}
which clearly vanish for odd values of $n$. Since the motion in the $y$ direction is purely diffusive, the moments are simply those of standard one-dimensional diffusion.

\section{Steady-state solution} \label{appendix_B}
For a particle starting its motion at $(0, y_0)$, Eq.~\eqref{eq:prop} and Eq.~\eqref{eq:phi_app} reduce to

\begin{equation}\label{eq:prop_app}
	p(x,y,t | 0, y_0) = \frac{\sqrt{3}}{2 \pi D t \sqrt{12 + (\dot \gamma t)^2}} e^{-\phi(x,y,t)},
\end{equation}
where 
\begin{align}
	\phi(x,y,t) & = \frac{\left(y-y_0\right)^2 \left(\dot\gamma^2 t^2+3\right)+3 (x-\dot\gamma t y)^2}{D t (12 + (\dot\gamma t)^2)} \nonumber \\
	& +\frac{3 \dot\gamma t \left(y - y_0\right) (x-\dot\gamma t y)}{D t (12 + (\dot\gamma t)^2)}.
\end{align}

We aim to derive the steady-state probability density under stochastic resetting using the renewal approach:

\begin{align}
	p_r(\bm{X}&,t|\bm{X}_0) = e^{- r t}p(\bm{X},t|\bm{X}_0) \\
	&+ r\int_0^t d\tau e^{-r\tau} \int d\bm{Y} p_r(\bm{Y},t-\tau|\bm{X}_0) p(\bm{X},\tau|\bm{X}_R),\nonumber
\end{align}
where $\bm{X} = (x,y)$ and $\bm{X}_R$ is the resetting location. The first term corresponds to trajectories where no resetting took place. The second term takes into account trajectories (with resetting) up to the time of the last resetting event before time $t$, i.e. at time $t-\tau$, when the particle is at position $\bm{Y}$. After the last reset, the particle propagates from the resetting location to $\bm{X}$ in the remaining time $\tau$. 

In the steady state, the renewal equation simplifies to

\begin{align}\label{renewal_ss}
	p_{ss}(\bm{X}|\bm{X}_0) &= r\int_0^\infty d\tau e^{-r\tau}  p(\bm{X},\tau|\bm{X}_R),
\end{align}
where \(p(\bm{X},\tau|\bm{X}_R)\) is given in Eq.\eqref{eq:prop} Obtaining an exact solution is challenging, so we use a perturbative approach by expanding \(\Phi(x, y, \tau)\) in powers of \(\dot\gamma\) up to the first order, given as


\begin{align}
	\phi(x, y, \tau) \approx \phi_0(x, y, \tau) + \dot{\gamma} \phi_1(x, y, \tau), 
\end{align}
where \(\phi_0(x, y, \tau)\) is the zeroth-order term (without shear, i.e., \(\dot{\gamma} = 0\)) and \(\phi_1(x, y, \tau)\) is the first-order correction due to shear. Therefore, the probability density becomes

\begin{align}
	p(x, y, \tau) \approx p_0(x, y, \tau) \left[ 1 - \dot{\gamma} \phi_1(x, y, \tau) \right],
\end{align}
where \(p_0(x, y, \tau)=\exp\left(-\phi_0(x, y, \tau)\right)/4\pi D\tau \) represents pure diffusion (without shear) and \(\phi_1(x, y, \tau)\) is the first-order correction, capturing the shear flow effect, given by

\begin{align} \label{phi_0}
	\phi_0(x, y, \tau) = \frac{(y - y_0)^2 + x^2}{4 D \tau}, 
\end{align}
and 
\begin{align}\label{phi_1}
	\phi_1(x, y) = \frac{x(y - y_0) - 2xy}{4 D}.
\end{align}
Using the renewal approach the steady-state solution is:

\begin{align}\label{perturbative}
	p_{ss}(x, y) \approx p_{ss}^{(0)}(x, y) - \dot{\gamma} p_{ss}^{(1)}(x, y),
\end{align}
where \(p_{ss}^{(0)}(x, y)\) and \(p_{ss}^{(1)}(x, y)\) are solutions to the renewal equation in Eq.\eqref{renewal_ss}. The zeroth-order (diffusion) term gives the well-known results\cite{evans2011diffusion}

\begin{align}
	p_{ss}^{(0)}(x, y) = \frac{r}{2 \pi D} K_0 \left( \alpha \sqrt{(y - y_0)^2 + x^2} \right),
\end{align}
where \(\alpha=\sqrt{r/D}\) and \(K_0\) is the modified Bessel function of the second kind (order zero). Similarly, we can calculate the first order correction, given by
\begin{equation}
	p_{ss}^{(1)}=r\int_{0}^{\infty}e^{-r\tau}p_0(x,y,\tau)\Phi_1(x, y)d\tau.
\end{equation}

Plugging \(p_0(x,y,\tau)\) and Eq.\eqref{phi_1} in the above equation and solving the integral gives
\begin{equation}
	p_{ss}^{(1)}(x,y,\tau)=\frac{r\left(x(y\!-\!y_0)\!-\!2xy\right)}{4D}K_0 \left( \alpha \sqrt{(y\!-\!y_0)^2\!+\!x^2} \right).
\end{equation}

Substituting these into Eq.\eqref{perturbative} yields the full perturbative solution

\begin{align}\label{eq_app:pdf_ss}
	p_{ss}(x, y) \approx \left( \frac{r}{2 \pi D} - \frac{\dot{\gamma} r \left[ x(y - y_0) - 2 x y \right]}{8 \pi D^2} \right) \nonumber \\
	\times K_0 \left( \alpha \sqrt{(y - y_0)^2 + x^2} \right).
\end{align}

We can calculate the probability fluxes in the system as

\begin{equation}
	\label{eq_app:fluxes}
	\mathbf{J}(x, y) = -D\nabla p_{ss}(x, y) + \mathbf{v}(x, y)p_{ss}(x, y),
\end{equation}
where $\mathbf{v}(x, y)=(\dot \gamma y, 0)$ is the drift velocity due to the shear flow, which acts in the $x$-direction and depends linearly on $y$. Using Eq.\eqref{eq_app:pdf_ss} the expressions for the fluxes read as

\begin{align}
	J_x(x, y) = & \frac{r}{8 D^2 \pi} \Bigg[ \dot{\gamma} \left( D(3y - y_0) + \dot{\gamma} x y (y + y_0) \right) \nonumber \\
	& \times K_0 \left( \alpha \sqrt{x^2 + (y - y_0)^2} \right) \nonumber \\
	& - (4D + \dot{\gamma} x (y + y_0)) D \nonumber \\
	& \times \left( -\alpha x \frac{K_1 \left( \alpha \sqrt{x^2 + (y - y_0)^2} \right)}{\sqrt{x^2 + (y - y_0)^2}} \right) \Bigg].
\end{align}

Similarly 

\begin{align}
	J_y(x, y) = & \frac{r}{8 D \pi} \Bigg[ -\dot{\gamma} x K_0 \left( \alpha \sqrt{x^2 + (y - y_0)^2} \right) \nonumber \\
	& + \alpha (y - y_0) \left( 4D + \dot{\gamma} x (y + y_0) \right) \nonumber \\
	& \times \frac{K_1 \left( \alpha \sqrt{x^2 + (y - y_0)^2} \right)}{\sqrt{x^2 + (y - y_0)^2}} \Bigg].
\end{align}
where  \(K_1\) is the modified Bessel function of the second kind of the first order.

\section{Moments under shear flow and a harmonic potential}\label{appendix_C}
To calculate the mean rate of thermodynamic work, we used the steady-state moments under the combined effect of shear flow and a harmonic potential. In this case, the Langevin equations can be written as
\begin{align}
	\dot x &=  -\lambda x + \dot\gamma y + \sqrt{2D}\xi_x(t),\\ 
	\dot y &=  -\lambda (y-y_R)+ \sqrt{2D}\xi_y(t),
\end{align}
where $\xi_x(t)$ and $\xi_y(t)$ are Gaussian white noises along the $i$ axis with $i=x, y$ with zero mean and Dirac delta time correlations $\langle \xi_i(t)\xi_j(t^\prime)\rangle = \delta_{ij}\delta(t-t^\prime)$.  Here we assumed that the potential is centered at $(0,y_R)$, and that its stiffness is $\lambda$. Since the $y$-direction is unaffected by the shear flow, we have the standard moments in harmonic potentials
\begin{align}
	\langle y(t) \rangle_\lambda &= y_R (1-e^{-\lambda t}),\\
	\langle y^2(t) \rangle_\lambda &=\frac{e^{-2 \lambda t}}{\lambda} (e^{\lambda t}-1) \left(D - y_R^2 \lambda  + e^{\lambda t}[D + y_R^2 \lambda]\right),
\end{align}
where we used the subscript $\lambda$ to denote averages in the presence of the harmonic potential. As a function of $y$, the motion in the $x$-direction can be obtained by explicitly solving the Langevin equation, e.g.
\begin{equation}
	x(t) = e^{-\lambda t}\int_0^t ds e^{\lambda s} \left(\dot \gamma y+ \sqrt{2D} \xi_x(s)\right).
\end{equation}
In the above we have ignored initial conditions as these do not matter in the steady state. From these results, we calculate the steady state moments explicitly:
\begin{align}
	\langle y\rangle_\lambda &=y_R,\\
	\langle y^2\rangle_\lambda &=y_R^2 +\frac{D}{\lambda},\\
	\langle x\rangle_\lambda &=\frac{\dot \gamma y_R}{\lambda},\\
	\langle x^2\rangle_\lambda &= \frac{D \lambda + y_R^2 \dot \gamma^2}{\lambda^2}.\\
\end{align}

\section*{REFERENCES}
%


\begin{thebibliography}{86}%
	\makeatletter
	\providecommand \@ifxundefined [1]{%
		\@ifx{#1\undefined}
	}%
	\providecommand \@ifnum [1]{%
		\ifnum #1\expandafter \@firstoftwo
		\else \expandafter \@secondoftwo
		\fi
	}%
	\providecommand \@ifx [1]{%
		\ifx #1\expandafter \@firstoftwo
		\else \expandafter \@secondoftwo
		\fi
	}%
	\providecommand \natexlab [1]{#1}%
	\providecommand \enquote  [1]{``#1''}%
	\providecommand \bibnamefont  [1]{#1}%
	\providecommand \bibfnamefont [1]{#1}%
	\providecommand \citenamefont [1]{#1}%
	\providecommand \href@noop [0]{\@secondoftwo}%
	\providecommand \href [0]{\begingroup \@sanitize@url \@href}%
	\providecommand \@href[1]{\@@startlink{#1}\@@href}%
	\providecommand \@@href[1]{\endgroup#1\@@endlink}%
	\providecommand \@sanitize@url [0]{\catcode `\\12\catcode `\$12\catcode
		`\&12\catcode `\#12\catcode `\^12\catcode `\_12\catcode `\%12\relax}%
	\providecommand \@@startlink[1]{}%
	\providecommand \@@endlink[0]{}%
	\providecommand \url  [0]{\begingroup\@sanitize@url \@url }%
	\providecommand \@url [1]{\endgroup\@href {#1}{\urlprefix }}%
	\providecommand \urlprefix  [0]{URL }%
	\providecommand \Eprint [0]{\href }%
	\providecommand \doibase [0]{http://dx.doi.org/}%
	\providecommand \selectlanguage [0]{\@gobble}%
	\providecommand \bibinfo  [0]{\@secondoftwo}%
	\providecommand \bibfield  [0]{\@secondoftwo}%
	\providecommand \translation [1]{[#1]}%
	\providecommand \BibitemOpen [0]{}%
	\providecommand \bibitemStop [0]{}%
	\providecommand \bibitemNoStop [0]{.\EOS\space}%
	\providecommand \EOS [0]{\spacefactor3000\relax}%
	\providecommand \BibitemShut  [1]{\csname bibitem#1\endcsname}%
	\let\auto@bib@innerbib\@empty
	\bibitem [{\citenamefont {Eckstein}, \citenamefont {Bailey},\ and\
		\citenamefont {Shapiro}(1977)}]{eckstein1977self}%
	\BibitemOpen
	\bibfield  {author} {\bibinfo {author} {\bibfnamefont {E.~C.}\ \bibnamefont
			{Eckstein}}, \bibinfo {author} {\bibfnamefont {D.~G.}\ \bibnamefont
			{Bailey}}, \ and\ \bibinfo {author} {\bibfnamefont {A.~H.}\ \bibnamefont
			{Shapiro}},\ }\bibfield  {title} {\enquote {\bibinfo {title} {Self-diffusion
				of particles in shear flow of a suspension},}\ }\href@noop {} {\bibfield
		{journal} {\bibinfo  {journal} {Journal of Fluid Mechanics}\ }\textbf
		{\bibinfo {volume} {79}},\ \bibinfo {pages} {191--208} (\bibinfo {year}
		{1977})}\BibitemShut {NoStop}%
	\bibitem [{\citenamefont {Morris}\ and\ \citenamefont
		{Brady}(1996)}]{morris1996self}%
	\BibitemOpen
	\bibfield  {author} {\bibinfo {author} {\bibfnamefont {J.~F.}\ \bibnamefont
			{Morris}}\ and\ \bibinfo {author} {\bibfnamefont {J.~F.}\ \bibnamefont
			{Brady}},\ }\bibfield  {title} {\enquote {\bibinfo {title} {Self-diffusion in
				sheared suspensions},}\ }\href@noop {} {\bibfield  {journal} {\bibinfo
			{journal} {Journal of Fluid Mechanics}\ }\textbf {\bibinfo {volume} {312}},\
		\bibinfo {pages} {223--252} (\bibinfo {year} {1996})}\BibitemShut {NoStop}%
	\bibitem [{\citenamefont {Frankel}\ and\ \citenamefont
		{Brenner}(1991)}]{frankel1991generalized}%
	\BibitemOpen
	\bibfield  {author} {\bibinfo {author} {\bibfnamefont {I.}~\bibnamefont
			{Frankel}}\ and\ \bibinfo {author} {\bibfnamefont {H.}~\bibnamefont
			{Brenner}},\ }\bibfield  {title} {\enquote {\bibinfo {title} {Generalized
				Taylor dispersion phenomena in unbounded homogeneous shear flows},}\
	}\href@noop {} {\bibfield  {journal} {\bibinfo  {journal} {Journal of Fluid
				Mechanics}\ }\textbf {\bibinfo {volume} {230}},\ \bibinfo {pages} {147--181}
		(\bibinfo {year} {1991})}\BibitemShut {NoStop}%
	\bibitem [{\citenamefont {San~Miguel}\ and\ \citenamefont
		{Sancho}(1979)}]{san1979brownian}%
	\BibitemOpen
	\bibfield  {author} {\bibinfo {author} {\bibfnamefont {M.}~\bibnamefont
			{San~Miguel}}\ and\ \bibinfo {author} {\bibfnamefont {J.}~\bibnamefont
			{Sancho}},\ }\bibfield  {title} {\enquote {\bibinfo {title} {Brownian motion
				in shear flow},}\ }\href@noop {} {\bibfield  {journal} {\bibinfo  {journal}
			{Physica A: Statistical Mechanics and its Applications}\ }\textbf {\bibinfo
			{volume} {99}},\ \bibinfo {pages} {357--364} (\bibinfo {year}
		{1979})}\BibitemShut {NoStop}%
	\bibitem [{\citenamefont {Novikov}(1958)}]{novikov1958concerning}%
	\BibitemOpen
	\bibfield  {author} {\bibinfo {author} {\bibfnamefont {E.}~\bibnamefont
			{Novikov}},\ }\bibfield  {title} {\enquote {\bibinfo {title} {Concerning a
				turbulent diffusion in a stream with a transverse gradient of velocity},}\
	}\href@noop {} {\bibfield  {journal} {\bibinfo  {journal} {Journal of Applied
				Mathematics and Mechanics}\ }\textbf {\bibinfo {volume} {22}},\ \bibinfo
		{pages} {576--579} (\bibinfo {year} {1958})}\BibitemShut {NoStop}%
	\bibitem [{\citenamefont {Makuch}\ \emph {et~al.}(2020)\citenamefont {Makuch},
		\citenamefont {Ho{\l}yst}, \citenamefont {Kalwarczyk}, \citenamefont
		{Garstecki},\ and\ \citenamefont {Brady}}]{makuch2020diffusion}%
	\BibitemOpen
	\bibfield  {author} {\bibinfo {author} {\bibfnamefont {K.}~\bibnamefont
			{Makuch}}, \bibinfo {author} {\bibfnamefont {R.}~\bibnamefont {Ho{\l}yst}},
		\bibinfo {author} {\bibfnamefont {T.}~\bibnamefont {Kalwarczyk}}, \bibinfo
		{author} {\bibfnamefont {P.}~\bibnamefont {Garstecki}}, \ and\ \bibinfo
		{author} {\bibfnamefont {J.~F.}\ \bibnamefont {Brady}},\ }\bibfield  {title}
	{\enquote {\bibinfo {title} {Diffusion and flow in complex liquids},}\
	}\href@noop {} {\bibfield  {journal} {\bibinfo  {journal} {Soft Matter}\
		}\textbf {\bibinfo {volume} {16}},\ \bibinfo {pages} {114--124} (\bibinfo
		{year} {2020})}\BibitemShut {NoStop}%
	\bibitem [{\citenamefont {Abdoli}\ \emph {et~al.}(2023)\citenamefont {Abdoli},
		\citenamefont {L{\"o}wen}, \citenamefont {Sommer},\ and\ \citenamefont
		{Sharma}}]{abdoli2023tailoring}%
	\BibitemOpen
	\bibfield  {author} {\bibinfo {author} {\bibfnamefont {I.}~\bibnamefont
			{Abdoli}}, \bibinfo {author} {\bibfnamefont {H.}~\bibnamefont {L{\"o}wen}},
		\bibinfo {author} {\bibfnamefont {J.-U.}\ \bibnamefont {Sommer}}, \ and\
		\bibinfo {author} {\bibfnamefont {A.}~\bibnamefont {Sharma}},\ }\bibfield
	{title} {\enquote {\bibinfo {title} {Tailoring the escape rate of a Brownian
				particle by combining a vortex flow with a magnetic field},}\ }\href@noop {}
	{\bibfield  {journal} {\bibinfo  {journal} {The Journal of Chemical Physics}\
		}\textbf {\bibinfo {volume} {158}} (\bibinfo {year} {2023})}\BibitemShut
	{NoStop}%
	\bibitem [{\citenamefont {T{\'o}thov{\'a}}\ and\ \citenamefont
		{Lis{\`y}}(2024)}]{tothova2024brownian}%
	\BibitemOpen
	\bibfield  {author} {\bibinfo {author} {\bibfnamefont {J.}~\bibnamefont
			{T{\'o}thov{\'a}}}\ and\ \bibinfo {author} {\bibfnamefont {V.}~\bibnamefont
			{Lis{\`y}}},\ }\bibfield  {title} {\enquote {\bibinfo {title} {Brownian
				motion in a viscous fluid of particles with constant and time-dependent
				friction},}\ }\href@noop {} {\bibfield  {journal} {\bibinfo  {journal}
			{Physics of Fluids}\ }\textbf {\bibinfo {volume} {36}} (\bibinfo {year}
		{2024})}\BibitemShut {NoStop}%
	\bibitem [{\citenamefont {Einstein}(1906)}]{einstein1906theorie}%
	\BibitemOpen
	\bibfield  {author} {\bibinfo {author} {\bibfnamefont {A.}~\bibnamefont
			{Einstein}},\ }\bibfield  {title} {\enquote {\bibinfo {title} {Zur Theorie
				der Brownschen Bewegung},}\ }\href@noop {} {\bibfield  {journal} {\bibinfo
			{journal} {Annalen der Physik}\ }\textbf {\bibinfo {volume} {324}},\ \bibinfo
		{pages} {371--381} (\bibinfo {year} {1906})}\BibitemShut {NoStop}%
	\bibitem [{\citenamefont {Von~Smoluchowski}(1906)}]{von1906kinetischen}%
	\BibitemOpen
	\bibfield  {author} {\bibinfo {author} {\bibfnamefont {M.}~\bibnamefont
			{Von~Smoluchowski}},\ }\bibfield  {title} {\enquote {\bibinfo {title} {Zur
				kinetischen Theorie der Brownschen Bolekularbewegung und der Suspensionen},}\
	}\href@noop {} {\bibfield  {journal} {\bibinfo  {journal} {Annalen der
				Physik}\ }\textbf {\bibinfo {volume} {326}},\ \bibinfo {pages} {756--780}
		(\bibinfo {year} {1906})}\BibitemShut {NoStop}%
	\bibitem [{\citenamefont {Taylor}(1953)}]{taylor1953dispersion}%
	\BibitemOpen
	\bibfield  {author} {\bibinfo {author} {\bibfnamefont {G.~I.}\ \bibnamefont
			{Taylor}},\ }\bibfield  {title} {\enquote {\bibinfo {title} {Dispersion of
				soluble matter in solvent flowing slowly through a tube},}\ }\href@noop {}
	{\bibfield  {journal} {\bibinfo  {journal} {Proceedings of the Royal Society
				of London. Series A. Mathematical and Physical Sciences}\ }\textbf {\bibinfo
			{volume} {219}},\ \bibinfo {pages} {186--203} (\bibinfo {year}
		{1953})}\BibitemShut {NoStop}%
	\bibitem [{\citenamefont {Aris}(1956)}]{aris1956dispersion}%
	\BibitemOpen
	\bibfield  {author} {\bibinfo {author} {\bibfnamefont {R.}~\bibnamefont
			{Aris}},\ }\bibfield  {title} {\enquote {\bibinfo {title} {On the dispersion
				of a solute in a fluid flowing through a tube},}\ }\href@noop {} {\bibfield
		{journal} {\bibinfo  {journal} {Proceedings of the Royal Society of London.
				Series A. Mathematical and Physical Sciences}\ }\textbf {\bibinfo {volume}
			{235}},\ \bibinfo {pages} {67--77} (\bibinfo {year} {1956})}\BibitemShut
	{NoStop}%
	\bibitem [{\citenamefont {Belongia}\ and\ \citenamefont
		{Baygents}(1997)}]{belongia1997measurements}%
	\BibitemOpen
	\bibfield  {author} {\bibinfo {author} {\bibfnamefont {B.}~\bibnamefont
			{Belongia}}\ and\ \bibinfo {author} {\bibfnamefont {J.}~\bibnamefont
			{Baygents}},\ }\bibfield  {title} {\enquote {\bibinfo {title} {Measurements
				on the diffusion coefficient of colloidal particles by Taylor--Aris
				dispersion},}\ }\href@noop {} {\bibfield  {journal} {\bibinfo  {journal}
			{Journal of Colloid and Interface Science}\ }\textbf {\bibinfo {volume}
			{195}},\ \bibinfo {pages} {19--31} (\bibinfo {year} {1997})}\BibitemShut
	{NoStop}%
	\bibitem [{\citenamefont {Huang}\ \emph {et~al.}(2011)\citenamefont {Huang},
		\citenamefont {Chavez}, \citenamefont {Taute}, \citenamefont {Luki{\'c}},
		\citenamefont {Jeney}, \citenamefont {Raizen},\ and\ \citenamefont
		{Florin}}]{huang2011direct}%
	\BibitemOpen
	\bibfield  {author} {\bibinfo {author} {\bibfnamefont {R.}~\bibnamefont
			{Huang}}, \bibinfo {author} {\bibfnamefont {I.}~\bibnamefont {Chavez}},
		\bibinfo {author} {\bibfnamefont {K.~M.}\ \bibnamefont {Taute}}, \bibinfo
		{author} {\bibfnamefont {B.}~\bibnamefont {Luki{\'c}}}, \bibinfo {author}
		{\bibfnamefont {S.}~\bibnamefont {Jeney}}, \bibinfo {author} {\bibfnamefont
			{M.~G.}\ \bibnamefont {Raizen}}, \ and\ \bibinfo {author} {\bibfnamefont
			{E.-L.}\ \bibnamefont {Florin}},\ }\bibfield  {title} {\enquote {\bibinfo
			{title} {Direct observation of the full transition from ballistic to
				diffusive Brownian motion in a liquid},}\ }\href@noop {} {\bibfield
		{journal} {\bibinfo  {journal} {Nature Physics}\ }\textbf {\bibinfo {volume}
			{7}},\ \bibinfo {pages} {576--580} (\bibinfo {year} {2011})}\BibitemShut
	{NoStop}%
	\bibitem [{\citenamefont {K{\"a}hlert}\ and\ \citenamefont
		{L{\"o}wen}(2012)}]{kahlert2012resonant}%
	\BibitemOpen
	\bibfield  {author} {\bibinfo {author} {\bibfnamefont {H.}~\bibnamefont
			{K{\"a}hlert}}\ and\ \bibinfo {author} {\bibfnamefont {H.}~\bibnamefont
			{L{\"o}wen}},\ }\bibfield  {title} {\enquote {\bibinfo {title} {Resonant
				behavior of trapped Brownian particles in an oscillatory shear flow},}\
	}\href@noop {} {\bibfield  {journal} {\bibinfo  {journal} {Physical Review
				E}\ }\textbf {\bibinfo
			{volume} {86}},\ \bibinfo {pages} {041119} (\bibinfo {year}
		{2012})}\BibitemShut {NoStop}%
	\bibitem [{\citenamefont {Kumar}\ \emph {et~al.}(2021)\citenamefont {Kumar},
		\citenamefont {Thomson}, \citenamefont {Powers},\ and\ \citenamefont
		{Harris}}]{kumar2021taylor}%
	\BibitemOpen
	\bibfield  {author} {\bibinfo {author} {\bibfnamefont {A.~H.}\ \bibnamefont
			{Kumar}}, \bibinfo {author} {\bibfnamefont {S.~J.}\ \bibnamefont {Thomson}},
		\bibinfo {author} {\bibfnamefont {T.~R.}\ \bibnamefont {Powers}}, \ and\
		\bibinfo {author} {\bibfnamefont {D.~M.}\ \bibnamefont {Harris}},\ }\bibfield
	{title} {\enquote {\bibinfo {title} {Taylor dispersion of elongated rods},}\
	}\href@noop {} {\bibfield  {journal} {\bibinfo  {journal} {Physical Review
				Fluids}\ }\textbf {\bibinfo {volume} {6}},\ \bibinfo {pages} {094501}
		(\bibinfo {year} {2021})}\BibitemShut {NoStop}%
	\bibitem [{\citenamefont {Ten~Hagen}, \citenamefont {Wittkowski},\ and\
		\citenamefont {L{\"o}wen}(2011)}]{ten2011brownian}%
	\BibitemOpen
	\bibfield  {author} {\bibinfo {author} {\bibfnamefont {B.}~\bibnamefont
			{Ten~Hagen}}, \bibinfo {author} {\bibfnamefont {R.}~\bibnamefont
			{Wittkowski}}, \ and\ \bibinfo {author} {\bibfnamefont {H.}~\bibnamefont
			{L{\"o}wen}},\ }\bibfield  {title} {\enquote {\bibinfo {title} {Brownian
				dynamics of a self-propelled particle in shear flow},}\ }\href@noop {}
	{\bibfield  {journal} {\bibinfo  {journal} {Physical Review E}\ }\textbf {\bibinfo {volume} {84}},\
		\bibinfo {pages} {031105} (\bibinfo {year} {2011})}\BibitemShut {NoStop}%
	\bibitem [{\citenamefont {Tarama}\ \emph {et~al.}(2013)\citenamefont {Tarama},
		\citenamefont {Menzel}, \citenamefont {Ten~Hagen}, \citenamefont
		{Wittkowski}, \citenamefont {Ohta},\ and\ \citenamefont
		{L{\"o}wen}}]{tarama2013dynamics}%
	\BibitemOpen
	\bibfield  {author} {\bibinfo {author} {\bibfnamefont {M.}~\bibnamefont
			{Tarama}}, \bibinfo {author} {\bibfnamefont {A.~M.}\ \bibnamefont {Menzel}},
		\bibinfo {author} {\bibfnamefont {B.}~\bibnamefont {Ten~Hagen}}, \bibinfo
		{author} {\bibfnamefont {R.}~\bibnamefont {Wittkowski}}, \bibinfo {author}
		{\bibfnamefont {T.}~\bibnamefont {Ohta}}, \ and\ \bibinfo {author}
		{\bibfnamefont {H.}~\bibnamefont {L{\"o}wen}},\ }\bibfield  {title} {\enquote
		{\bibinfo {title} {Dynamics of a deformable active particle under shear
				flow},}\ }\href@noop {} {\bibfield  {journal} {\bibinfo  {journal} {The
				Journal of Chemical Physics}\ }\textbf {\bibinfo {volume} {139}} (\bibinfo
		{year} {2013})}\BibitemShut {NoStop}%
	\bibitem [{\citenamefont {Sandoval}, \citenamefont {Hidalgo-Gonzalez},\ and\
		\citenamefont {Jimenez-Aquino}(2018)}]{sandoval2018self}%
	\BibitemOpen
	\bibfield  {author} {\bibinfo {author} {\bibfnamefont {M.}~\bibnamefont
			{Sandoval}}, \bibinfo {author} {\bibfnamefont {J.~C.}\ \bibnamefont
			{Hidalgo-Gonzalez}}, \ and\ \bibinfo {author} {\bibfnamefont {J.~I.}\
			\bibnamefont {Jimenez-Aquino}},\ }\bibfield  {title} {\enquote {\bibinfo
			{title} {Self-driven particles in linear flows and trapped in a harmonic
				potential},}\ }\href@noop {} {\bibfield  {journal} {\bibinfo  {journal}
			{Physical Review E}\ }\textbf {\bibinfo {volume} {97}},\ \bibinfo {pages}
		{032603} (\bibinfo {year} {2018})}\BibitemShut {NoStop}%
	\bibitem [{\citenamefont {Asheichyk}, \citenamefont {Fuchs},\ and\
		\citenamefont {Kr{\"u}ger}(2021)}]{asheichyk2021brownian}%
	\BibitemOpen
	\bibfield  {author} {\bibinfo {author} {\bibfnamefont {K.}~\bibnamefont
			{Asheichyk}}, \bibinfo {author} {\bibfnamefont {M.}~\bibnamefont {Fuchs}}, \
		and\ \bibinfo {author} {\bibfnamefont {M.}~\bibnamefont {Kr{\"u}ger}},\
	}\bibfield  {title} {\enquote {\bibinfo {title} {Brownian systems perturbed
				by mild shear: comparing response relations},}\ }\href@noop {} {\bibfield
		{journal} {\bibinfo  {journal} {Journal of Physics: Condensed Matter}\
		}\textbf {\bibinfo {volume} {33}},\ \bibinfo {pages} {405101} (\bibinfo
		{year} {2021})}\BibitemShut {NoStop}%
	\bibitem [{\citenamefont {Ziehl}\ \emph {et~al.}(2009)\citenamefont {Ziehl},
		\citenamefont {Bammert}, \citenamefont {Holzer}, \citenamefont {Wagner},\
		and\ \citenamefont {Zimmermann}}]{ziehl2009direct}%
	\BibitemOpen
	\bibfield  {author} {\bibinfo {author} {\bibfnamefont {A.}~\bibnamefont
			{Ziehl}}, \bibinfo {author} {\bibfnamefont {J.}~\bibnamefont {Bammert}},
		\bibinfo {author} {\bibfnamefont {L.}~\bibnamefont {Holzer}}, \bibinfo
		{author} {\bibfnamefont {C.}~\bibnamefont {Wagner}}, \ and\ \bibinfo {author}
		{\bibfnamefont {W.}~\bibnamefont {Zimmermann}},\ }\bibfield  {title}
	{\enquote {\bibinfo {title} {Direct measurement of shear-induced
				cross-correlations of Brownian motion},}\ }\href@noop {} {\bibfield
		{journal} {\bibinfo  {journal} {Physical Review Letters}\ }\textbf {\bibinfo
			{volume} {103}},\ \bibinfo {pages} {230602} (\bibinfo {year}
		{2009})}\BibitemShut {NoStop}%
	\bibitem [{\citenamefont {Howard}\ \emph {et~al.}(2016)\citenamefont {Howard},
		\citenamefont {Gautam}, \citenamefont {Panagiotopoulos},\ and\ \citenamefont
		{Nikoubashman}}]{howard2016axial}%
	\BibitemOpen
	\bibfield  {author} {\bibinfo {author} {\bibfnamefont {M.~P.}\ \bibnamefont
			{Howard}}, \bibinfo {author} {\bibfnamefont {A.}~\bibnamefont {Gautam}},
		\bibinfo {author} {\bibfnamefont {A.~Z.}\ \bibnamefont {Panagiotopoulos}}, \
		and\ \bibinfo {author} {\bibfnamefont {A.}~\bibnamefont {Nikoubashman}},\
	}\bibfield  {title} {\enquote {\bibinfo {title} {Axial dispersion of Brownian
				colloids in microfluidic channels},}\ }\href@noop {} {\bibfield  {journal}
		{\bibinfo  {journal} {Physical Review Fluids}\ }\textbf {\bibinfo {volume}
			{1}},\ \bibinfo {pages} {044203} (\bibinfo {year} {2016})}\BibitemShut
	{NoStop}%
	\bibitem [{\citenamefont {Holzer}\ \emph {et~al.}(2010)\citenamefont {Holzer},
		\citenamefont {Bammert}, \citenamefont {Rzehak},\ and\ \citenamefont
		{Zimmermann}}]{holzer2010dynamics}%
	\BibitemOpen
	\bibfield  {author} {\bibinfo {author} {\bibfnamefont {L.}~\bibnamefont
			{Holzer}}, \bibinfo {author} {\bibfnamefont {J.}~\bibnamefont {Bammert}},
		\bibinfo {author} {\bibfnamefont {R.}~\bibnamefont {Rzehak}}, \ and\ \bibinfo
		{author} {\bibfnamefont {W.}~\bibnamefont {Zimmermann}},\ }\bibfield  {title}
	{\enquote {\bibinfo {title} {Dynamics of a trapped Brownian particle in shear
				flows},}\ }\href@noop {} {\bibfield  {journal} {\bibinfo  {journal} {Physical
				Review E}\ }\textbf
		{\bibinfo {volume} {81}},\ \bibinfo {pages} {041124} (\bibinfo {year}
		{2010})}\BibitemShut {NoStop}%
	\bibitem [{\citenamefont {Bammert}\ and\ \citenamefont
		{Zimmermann}(2010)}]{bammert2010probability}%
	\BibitemOpen
	\bibfield  {author} {\bibinfo {author} {\bibfnamefont {J.}~\bibnamefont
			{Bammert}}\ and\ \bibinfo {author} {\bibfnamefont {W.}~\bibnamefont
			{Zimmermann}},\ }\bibfield  {title} {\enquote {\bibinfo {title} {Probability
				distribution of a trapped Brownian particle in plane shear flows},}\
	}\href@noop {} {\bibfield  {journal} {\bibinfo  {journal} {Physical Review
				E}\ }\textbf {\bibinfo
			{volume} {82}},\ \bibinfo {pages} {052102} (\bibinfo {year}
		{2010})}\BibitemShut {NoStop}%
	\bibitem [{\citenamefont {Orihara}\ and\ \citenamefont
		{Takikawa}(2011)}]{orihara2011brownian}%
	\BibitemOpen
	\bibfield  {author} {\bibinfo {author} {\bibfnamefont {H.}~\bibnamefont
			{Orihara}}\ and\ \bibinfo {author} {\bibfnamefont {Y.}~\bibnamefont
			{Takikawa}},\ }\bibfield  {title} {\enquote {\bibinfo {title} {Brownian
				motion in shear flow: Direct observation of anomalous diffusion},}\
	}\href@noop {} {\bibfield  {journal} {\bibinfo  {journal} {Physical Review
				E}\ }\textbf {\bibinfo
			{volume} {84}},\ \bibinfo {pages} {061120} (\bibinfo {year}
		{2011})}\BibitemShut {NoStop}%
	\bibitem [{\citenamefont {Evans}\ and\ \citenamefont
		{Majumdar}(2011{\natexlab{a}})}]{evans2011diffusion}%
	\BibitemOpen
	\bibfield  {author} {\bibinfo {author} {\bibfnamefont {M.~R.}\ \bibnamefont
			{Evans}}\ and\ \bibinfo {author} {\bibfnamefont {S.~N.}\ \bibnamefont
			{Majumdar}},\ }\bibfield  {title} {\enquote {\bibinfo {title} {Diffusion with
				stochastic resetting},}\ }\href@noop {} {\bibfield  {journal} {\bibinfo
			{journal} {Physical Review Letters}\ }\textbf {\bibinfo {volume} {106}},\
		\bibinfo {pages} {160601} (\bibinfo {year} {2011}{\natexlab{a}})}\BibitemShut
	{NoStop}%
	\bibitem [{\citenamefont {Evans}\ and\ \citenamefont
		{Majumdar}(2011{\natexlab{b}})}]{evans2011optimal}%
	\BibitemOpen
	\bibfield  {author} {\bibinfo {author} {\bibfnamefont {M.~R.}\ \bibnamefont
			{Evans}}\ and\ \bibinfo {author} {\bibfnamefont {S.~N.}\ \bibnamefont
			{Majumdar}},\ }\bibfield  {title} {\enquote {\bibinfo {title} {Diffusion with
				optimal resetting},}\ }\href {\doibase 10.1088/1751-8113/44/43/435001}
	{\bibfield  {journal} {\bibinfo  {journal} {Journal of Physics A: Mathematical and Theoretical}\
		}\textbf {\bibinfo {volume} {44}},\ \bibinfo {pages} {435001} (\bibinfo
		{year} {2011}{\natexlab{b}})}\BibitemShut {NoStop}%
	\bibitem [{\citenamefont {Evans}, \citenamefont {Majumdar},\ and\ \citenamefont
		{Schehr}(2020)}]{evans2020review}%
	\BibitemOpen
	\bibfield  {author} {\bibinfo {author} {\bibfnamefont {M.~R.}\ \bibnamefont
			{Evans}}, \bibinfo {author} {\bibfnamefont {S.~N.}\ \bibnamefont {Majumdar}},
		\ and\ \bibinfo {author} {\bibfnamefont {G.}~\bibnamefont {Schehr}},\
	}\bibfield  {title} {\enquote {\bibinfo {title} {Stochastic resetting and
				applications},}\ }\href@noop {} {\bibfield  {journal} {\bibinfo  {journal}
			{Journal of Physics A: Mathematical and Theoretical}\ }\textbf {\bibinfo {volume} {53}},\ \bibinfo
		{pages} {193001} (\bibinfo {year} {2020})}\BibitemShut {NoStop}%
	\bibitem [{\citenamefont {Majumdar}, \citenamefont {Sabhapandit},\ and\
		\citenamefont {Schehr}(2015)}]{majumdar2015dynamical}%
	\BibitemOpen
	\bibfield  {author} {\bibinfo {author} {\bibfnamefont {S.~N.}\ \bibnamefont
			{Majumdar}}, \bibinfo {author} {\bibfnamefont {S.}~\bibnamefont
			{Sabhapandit}}, \ and\ \bibinfo {author} {\bibfnamefont {G.}~\bibnamefont
			{Schehr}},\ }\bibfield  {title} {\enquote {\bibinfo {title} {Dynamical
				transition in the temporal relaxation of stochastic processes under
				resetting},}\ }\href@noop {} {\bibfield  {journal} {\bibinfo  {journal}
			{Physical Review E}\ }\textbf {\bibinfo {volume} {91}},\ \bibinfo {pages}
		{052131} (\bibinfo {year} {2015})}\BibitemShut {NoStop}%
	\bibitem [{\citenamefont {Santra}, \citenamefont {Basu},\ and\ \citenamefont
		{Sabhapandit}(2020)}]{santra2020run}%
	\BibitemOpen
	\bibfield  {author} {\bibinfo {author} {\bibfnamefont {I.}~\bibnamefont
			{Santra}}, \bibinfo {author} {\bibfnamefont {U.}~\bibnamefont {Basu}}, \ and\
		\bibinfo {author} {\bibfnamefont {S.}~\bibnamefont {Sabhapandit}},\
	}\bibfield  {title} {\enquote {\bibinfo {title} {Run-and-tumble particles in
				two dimensions under stochastic resetting conditions},}\ }\href@noop {}
	{\bibfield  {journal} {\bibinfo  {journal} {Journal of Statistical Mechanics:
				Theory and Experiment}\ }\textbf {\bibinfo {volume} {2020}},\ \bibinfo
		{pages} {113206} (\bibinfo {year} {2020})}\BibitemShut {NoStop}%
	\bibitem [{\citenamefont {Ku{\'s}mierz}\ and\ \citenamefont
		{Gudowska-Nowak}(2015)}]{kusmierz2015optimal}%
	\BibitemOpen
	\bibfield  {author} {\bibinfo {author} {\bibfnamefont {{\L}.}~\bibnamefont
			{Ku{\'s}mierz}}\ and\ \bibinfo {author} {\bibfnamefont {E.}~\bibnamefont
			{Gudowska-Nowak}},\ }\bibfield  {title} {\enquote {\bibinfo {title} {Optimal
				first-arrival times in L{\'e}vy flights with resetting},}\ }\href@noop {}
	{\bibfield  {journal} {\bibinfo  {journal} {Physical Review E}\ }\textbf
		{\bibinfo {volume} {92}},\ \bibinfo {pages} {052127} (\bibinfo {year}
		{2015})}\BibitemShut {NoStop}%
	\bibitem [{\citenamefont {Reuveni}(2016)}]{reuveni2016optimal}%
	\BibitemOpen
	\bibfield  {author} {\bibinfo {author} {\bibfnamefont {S.}~\bibnamefont
			{Reuveni}},\ }\bibfield  {title} {\enquote {\bibinfo {title} {Optimal
				stochastic restart renders fluctuations in first passage times universal},}\
	}\href@noop {} {\bibfield  {journal} {\bibinfo  {journal} {Physical Review
				Letters}\ }\textbf {\bibinfo {volume} {116}},\ \bibinfo {pages} {170601}
		(\bibinfo {year} {2016})}\BibitemShut {NoStop}%
	\bibitem [{\citenamefont {Evans}\ and\ \citenamefont
		{Ray}(2024)}]{evans2024stochasticresettingprevailssharp}%
	\BibitemOpen
	\bibfield  {author} {\bibinfo {author} {\bibfnamefont {M.~R.}\ \bibnamefont
			{Evans}}\ and\ \bibinfo {author} {\bibfnamefont {S.}~\bibnamefont {Ray}},\
	}\href {https://arxiv.org/abs/2410.01941} {\enquote {\bibinfo {title}
			{Stochastic resetting prevails over sharp restart for broad target
				distributions},}\ } (\bibinfo {year} {2024}),\ \Eprint
	{http://arxiv.org/abs/2410.01941} {arXiv:2410.01941 [cond-mat.stat-mech]}
	\BibitemShut {NoStop}%
	\bibitem [{\citenamefont {Rotbart}, \citenamefont {Reuveni},\ and\
		\citenamefont {Urbakh}(2015)}]{rotbart2015michaelis}%
	\BibitemOpen
	\bibfield  {author} {\bibinfo {author} {\bibfnamefont {T.}~\bibnamefont
			{Rotbart}}, \bibinfo {author} {\bibfnamefont {S.}~\bibnamefont {Reuveni}}, \
		and\ \bibinfo {author} {\bibfnamefont {M.}~\bibnamefont {Urbakh}},\
	}\bibfield  {title} {\enquote {\bibinfo {title} {Michaelis-Menten reaction
				scheme as a unified approach towards the optimal restart problem},}\
	}\href@noop {} {\bibfield  {journal} {\bibinfo  {journal} {Physical Review
				E}\ }\textbf {\bibinfo {volume} {92}},\ \bibinfo {pages} {060101} (\bibinfo
		{year} {2015})}\BibitemShut {NoStop}%
	\bibitem [{\citenamefont {Sandev}\ \emph
		{et~al.}(2022{\natexlab{a}})\citenamefont {Sandev}, \citenamefont
		{Domazetoski}, \citenamefont {Kocarev}, \citenamefont {Metzler},\ and\
		\citenamefont {Chechkin}}]{sandev2022heterogeneous}%
	\BibitemOpen
	\bibfield  {author} {\bibinfo {author} {\bibfnamefont {T.}~\bibnamefont
			{Sandev}}, \bibinfo {author} {\bibfnamefont {V.}~\bibnamefont {Domazetoski}},
		\bibinfo {author} {\bibfnamefont {L.}~\bibnamefont {Kocarev}}, \bibinfo
		{author} {\bibfnamefont {R.}~\bibnamefont {Metzler}}, \ and\ \bibinfo
		{author} {\bibfnamefont {A.}~\bibnamefont {Chechkin}},\ }\bibfield  {title}
	{\enquote {\bibinfo {title} {Heterogeneous diffusion with stochastic
				resetting},}\ }\href@noop {} {\bibfield  {journal} {\bibinfo  {journal}
			{Journal of Physics A: Mathematical and Theoretical}\ }\textbf {\bibinfo
			{volume} {55}},\ \bibinfo {pages} {074003} (\bibinfo {year}
		{2022}{\natexlab{a}})}\BibitemShut {NoStop}%
	\bibitem [{\citenamefont {Wang}\ \emph {et~al.}(2021)\citenamefont {Wang},
		\citenamefont {Cherstvy}, \citenamefont {Kantz}, \citenamefont {Metzler},\
		and\ \citenamefont {Sokolov}}]{wang2021time}%
	\BibitemOpen
	\bibfield  {author} {\bibinfo {author} {\bibfnamefont {W.}~\bibnamefont
			{Wang}}, \bibinfo {author} {\bibfnamefont {A.~G.}\ \bibnamefont {Cherstvy}},
		\bibinfo {author} {\bibfnamefont {H.}~\bibnamefont {Kantz}}, \bibinfo
		{author} {\bibfnamefont {R.}~\bibnamefont {Metzler}}, \ and\ \bibinfo
		{author} {\bibfnamefont {I.~M.}\ \bibnamefont {Sokolov}},\ }\bibfield
	{title} {\enquote {\bibinfo {title} {Time averaging and emerging
				nonergodicity upon resetting of fractional Brownian motion and heterogeneous
				diffusion processes},}\ }\href@noop {} {\bibfield  {journal} {\bibinfo
			{journal} {Physical Review E}\ }\textbf {\bibinfo {volume} {104}},\ \bibinfo
		{pages} {024105} (\bibinfo {year} {2021})}\BibitemShut {NoStop}%
	\bibitem [{\citenamefont {Ray}(2020)}]{ray2020space}%
	\BibitemOpen
	\bibfield  {author} {\bibinfo {author} {\bibfnamefont {S.}~\bibnamefont
			{Ray}},\ }\bibfield  {title} {\enquote {\bibinfo {title} {Space-dependent
				diffusion with stochastic resetting: A first-passage study},}\ }\href@noop {}
	{\bibfield  {journal} {\bibinfo  {journal} {The Journal of Chemical Physics}\
		}\textbf {\bibinfo {volume} {153}},\ \bibinfo {pages} {234904} (\bibinfo
		{year} {2020})}\BibitemShut {NoStop}%
	\bibitem [{\citenamefont {Mutothya}\ \emph {et~al.}(2021)\citenamefont
		{Mutothya}, \citenamefont {Xu}, \citenamefont {Li}, \citenamefont {Metzler},\
		and\ \citenamefont {Mutua}}]{mutothya2021first}%
	\BibitemOpen
	\bibfield  {author} {\bibinfo {author} {\bibfnamefont {N.~M.}\ \bibnamefont
			{Mutothya}}, \bibinfo {author} {\bibfnamefont {Y.}~\bibnamefont {Xu}},
		\bibinfo {author} {\bibfnamefont {Y.}~\bibnamefont {Li}}, \bibinfo {author}
		{\bibfnamefont {R.}~\bibnamefont {Metzler}}, \ and\ \bibinfo {author}
		{\bibfnamefont {N.~M.}\ \bibnamefont {Mutua}},\ }\bibfield  {title} {\enquote
		{\bibinfo {title} {First passage dynamics of stochastic motion in
				heterogeneous media driven by correlated white Gaussian and coloured
				non-Gaussian noises},}\ }\href@noop {} {\bibfield  {journal} {\bibinfo
			{journal} {Journal of Physics: Complexity}\ }\textbf {\bibinfo {volume}
			{2}},\ \bibinfo {pages} {045012} (\bibinfo {year} {2021})}\BibitemShut
	{NoStop}%
	\bibitem [{\citenamefont {Lenzi}\ \emph {et~al.}(2022)\citenamefont {Lenzi},
		\citenamefont {Lenzi}, \citenamefont {Guilherme}, \citenamefont
		{Evangelista},\ and\ \citenamefont {Ribeiro}}]{lenzi2022transient}%
	\BibitemOpen
	\bibfield  {author} {\bibinfo {author} {\bibfnamefont {M.}~\bibnamefont
			{Lenzi}}, \bibinfo {author} {\bibfnamefont {E.}~\bibnamefont {Lenzi}},
		\bibinfo {author} {\bibfnamefont {L.}~\bibnamefont {Guilherme}}, \bibinfo
		{author} {\bibfnamefont {L.}~\bibnamefont {Evangelista}}, \ and\ \bibinfo
		{author} {\bibfnamefont {H.}~\bibnamefont {Ribeiro}},\ }\bibfield  {title}
	{\enquote {\bibinfo {title} {Transient anomalous diffusion in heterogeneous
				media with stochastic resetting},}\ }\href@noop {} {\bibfield  {journal}
		{\bibinfo  {journal} {Physica A: Statistical Mechanics and its Applications}\
		}\textbf {\bibinfo {volume} {588}},\ \bibinfo {pages} {126560} (\bibinfo
		{year} {2022})}\BibitemShut {NoStop}%
	\bibitem [{\citenamefont {Sandev}\ and\ \citenamefont
		{Iomin}(2024)}]{sandev2024fractional}%
	\BibitemOpen
	\bibfield  {author} {\bibinfo {author} {\bibfnamefont {T.}~\bibnamefont
			{Sandev}}\ and\ \bibinfo {author} {\bibfnamefont {A.}~\bibnamefont {Iomin}},\
	}\bibfield  {title} {\enquote {\bibinfo {title} {Fractional heterogeneous
				telegraph processes: Interplay between heterogeneity, memory, and stochastic
				resetting},}\ }\href@noop {} {\bibfield  {journal} {\bibinfo  {journal}
			{Physical Review E}\ }\textbf {\bibinfo {volume} {110}},\ \bibinfo {pages}
		{024101} (\bibinfo {year} {2024})}\BibitemShut {NoStop}%
	\bibitem [{\citenamefont {Menon~Jr}\ and\ \citenamefont
		{Anteneodo}(2024)}]{menon2024random}%
	\BibitemOpen
	\bibfield  {author} {\bibinfo {author} {\bibfnamefont {L.}~\bibnamefont
			{Menon~Jr}}\ and\ \bibinfo {author} {\bibfnamefont {C.}~\bibnamefont
			{Anteneodo}},\ }\bibfield  {title} {\enquote {\bibinfo {title} {Random search
				with resetting in heterogeneous environments},}\ }\href@noop {} {\bibfield
		{journal} {\bibinfo  {journal} {arXiv preprint arXiv:2408.04726}\ } (\bibinfo
		{year} {2024})}\BibitemShut {NoStop}%
	\bibitem [{\citenamefont {Sandev}\ \emph
		{et~al.}(2022{\natexlab{b}})\citenamefont {Sandev}, \citenamefont {Kocarev},
		\citenamefont {Metzler},\ and\ \citenamefont
		{Chechkin}}]{sandev2022stochastic}%
	\BibitemOpen
	\bibfield  {author} {\bibinfo {author} {\bibfnamefont {T.}~\bibnamefont
			{Sandev}}, \bibinfo {author} {\bibfnamefont {L.}~\bibnamefont {Kocarev}},
		\bibinfo {author} {\bibfnamefont {R.}~\bibnamefont {Metzler}}, \ and\
		\bibinfo {author} {\bibfnamefont {A.}~\bibnamefont {Chechkin}},\ }\bibfield
	{title} {\enquote {\bibinfo {title} {Stochastic dynamics with multiplicative
				dichotomic noise: Heterogeneous telegrapher’s equation, anomalous
				crossovers and resetting},}\ }\href@noop {} {\bibfield  {journal} {\bibinfo
			{journal} {Chaos, Solitons \& Fractals}\ }\textbf {\bibinfo {volume} {165}},\
		\bibinfo {pages} {112878} (\bibinfo {year} {2022}{\natexlab{b}})}\BibitemShut
	{NoStop}%
	\bibitem [{\citenamefont {Bressloff}(2020)}]{bressloff2020switching}%
	\BibitemOpen
	\bibfield  {author} {\bibinfo {author} {\bibfnamefont {P.~C.}\ \bibnamefont
			{Bressloff}},\ }\bibfield  {title} {\enquote {\bibinfo {title} {Switching
				diffusions and stochastic resetting},}\ }\href@noop {} {\bibfield  {journal}
		{\bibinfo  {journal} {Journal of Physics A: Mathematical and Theoretical}\
		}\textbf {\bibinfo {volume} {53}},\ \bibinfo {pages} {275003} (\bibinfo
		{year} {2020})}\BibitemShut {NoStop}%
	\bibitem [{\citenamefont {Abdoli}\ \emph {et~al.}(2020)\citenamefont {Abdoli},
		\citenamefont {Vuijk}, \citenamefont {Wittmann}, \citenamefont {Sommer},
		\citenamefont {Brader},\ and\ \citenamefont {Sharma}}]{abdoli2020stationary}%
	\BibitemOpen
	\bibfield  {author} {\bibinfo {author} {\bibfnamefont {I.}~\bibnamefont
			{Abdoli}}, \bibinfo {author} {\bibfnamefont {H.~D.}\ \bibnamefont {Vuijk}},
		\bibinfo {author} {\bibfnamefont {R.}~\bibnamefont {Wittmann}}, \bibinfo
		{author} {\bibfnamefont {J.-U.}\ \bibnamefont {Sommer}}, \bibinfo {author}
		{\bibfnamefont {J.~M.}\ \bibnamefont {Brader}}, \ and\ \bibinfo {author}
		{\bibfnamefont {A.}~\bibnamefont {Sharma}},\ }\bibfield  {title} {\enquote
		{\bibinfo {title} {Stationary state in Brownian systems with Lorentz
				force},}\ }\href@noop {} {\bibfield  {journal} {\bibinfo  {journal} {Physical
				Review Research}\ }\textbf {\bibinfo {volume} {2}},\ \bibinfo {pages}
		{023381} (\bibinfo {year} {2020})}\BibitemShut {NoStop}%
	\bibitem [{\citenamefont {Abdoli}\ and\ \citenamefont
		{Sharma}(2021)}]{abdoli2021stochastic}%
	\BibitemOpen
	\bibfield  {author} {\bibinfo {author} {\bibfnamefont {I.}~\bibnamefont
			{Abdoli}}\ and\ \bibinfo {author} {\bibfnamefont {A.}~\bibnamefont
			{Sharma}},\ }\bibfield  {title} {\enquote {\bibinfo {title} {Stochastic
				resetting of active Brownian particles with Lorentz force},}\ }\href@noop {}
	{\bibfield  {journal} {\bibinfo  {journal} {Soft Matter}\ }\textbf {\bibinfo
			{volume} {17}},\ \bibinfo {pages} {1307--1316} (\bibinfo {year}
		{2021})}\BibitemShut {NoStop}%
	\bibitem [{\citenamefont {Domazetoski}\ \emph {et~al.}(2020)\citenamefont
		{Domazetoski}, \citenamefont {Mas{\'o}-Puigdellosas}, \citenamefont {Sandev},
		\citenamefont {M{\'e}ndez}, \citenamefont {Iomin},\ and\ \citenamefont
		{Kocarev}}]{domazetoski2020stochastic}%
	\BibitemOpen
	\bibfield  {author} {\bibinfo {author} {\bibfnamefont {V.}~\bibnamefont
			{Domazetoski}}, \bibinfo {author} {\bibfnamefont {A.}~\bibnamefont
			{Mas{\'o}-Puigdellosas}}, \bibinfo {author} {\bibfnamefont {T.}~\bibnamefont
			{Sandev}}, \bibinfo {author} {\bibfnamefont {V.}~\bibnamefont {M{\'e}ndez}},
		\bibinfo {author} {\bibfnamefont {A.}~\bibnamefont {Iomin}}, \ and\ \bibinfo
		{author} {\bibfnamefont {L.}~\bibnamefont {Kocarev}},\ }\bibfield  {title}
	{\enquote {\bibinfo {title} {Stochastic resetting on comblike structures},}\
	}\href@noop {} {\bibfield  {journal} {\bibinfo  {journal} {Physical Review
				Research}\ }\textbf {\bibinfo {volume} {2}},\ \bibinfo {pages} {033027}
		(\bibinfo {year} {2020})}\BibitemShut {NoStop}%
	\bibitem [{\citenamefont {Antonio Faustino~dos
			Santos}(2020)}]{antonio2020comb}%
	\BibitemOpen
	\bibfield  {author} {\bibinfo {author} {\bibfnamefont {M.}~\bibnamefont
			{Antonio Faustino~dos Santos}},\ }\bibfield  {title} {\enquote {\bibinfo
			{title} {Comb model with non-static stochastic resetting and anomalous
				diffusion},}\ }\href@noop {} {\bibfield  {journal} {\bibinfo  {journal}
			{Fractal and Fractional}\ }\textbf {\bibinfo {volume} {4}},\ \bibinfo {pages}
		{28} (\bibinfo {year} {2020})}\BibitemShut {NoStop}%
	\bibitem [{\citenamefont {Mas{\'o}-Puigdellosas}, \citenamefont {Sandev},\ and\
		\citenamefont {M{\'e}ndez}(2023)}]{maso2023random}%
	\BibitemOpen
	\bibfield  {author} {\bibinfo {author} {\bibfnamefont {A.}~\bibnamefont
			{Mas{\'o}-Puigdellosas}}, \bibinfo {author} {\bibfnamefont {T.}~\bibnamefont
			{Sandev}}, \ and\ \bibinfo {author} {\bibfnamefont {V.}~\bibnamefont
			{M{\'e}ndez}},\ }\bibfield  {title} {\enquote {\bibinfo {title} {Random walks
				on comb-like structures under stochastic resetting},}\ }\href@noop {}
	{\bibfield  {journal} {\bibinfo  {journal} {Entropy}\ }\textbf {\bibinfo
			{volume} {25}},\ \bibinfo {pages} {1529} (\bibinfo {year}
		{2023})}\BibitemShut {NoStop}%
	\bibitem [{\citenamefont {Singh}\ \emph {et~al.}(2021)\citenamefont {Singh},
		\citenamefont {Sandev}, \citenamefont {Iomin},\ and\ \citenamefont
		{Metzler}}]{singh2021backbone}%
	\BibitemOpen
	\bibfield  {author} {\bibinfo {author} {\bibfnamefont {R.}~\bibnamefont
			{Singh}}, \bibinfo {author} {\bibfnamefont {T.}~\bibnamefont {Sandev}},
		\bibinfo {author} {\bibfnamefont {A.}~\bibnamefont {Iomin}}, \ and\ \bibinfo
		{author} {\bibfnamefont {R.}~\bibnamefont {Metzler}},\ }\bibfield  {title}
	{\enquote {\bibinfo {title} {Backbone diffusion and first-passage dynamics in
				a comb structure with confining branches under stochastic resetting},}\
	}\href@noop {} {\bibfield  {journal} {\bibinfo  {journal} {Journal of Physics
				A: Mathematical and Theoretical}\ }\textbf {\bibinfo {volume} {54}},\
		\bibinfo {pages} {404006} (\bibinfo {year} {2021})}\BibitemShut {NoStop}%
	\bibitem [{\citenamefont {Trajanovski}\ \emph {et~al.}(2023)\citenamefont
		{Trajanovski}, \citenamefont {Jolakoski}, \citenamefont {Zelenkovski},
		\citenamefont {Iomin}, \citenamefont {Kocarev},\ and\ \citenamefont
		{Sandev}}]{trajanovski2023ornstein}%
	\BibitemOpen
	\bibfield  {author} {\bibinfo {author} {\bibfnamefont {P.}~\bibnamefont
			{Trajanovski}}, \bibinfo {author} {\bibfnamefont {P.}~\bibnamefont
			{Jolakoski}}, \bibinfo {author} {\bibfnamefont {K.}~\bibnamefont
			{Zelenkovski}}, \bibinfo {author} {\bibfnamefont {A.}~\bibnamefont {Iomin}},
		\bibinfo {author} {\bibfnamefont {L.}~\bibnamefont {Kocarev}}, \ and\
		\bibinfo {author} {\bibfnamefont {T.}~\bibnamefont {Sandev}},\ }\bibfield
	{title} {\enquote {\bibinfo {title} {Ornstein-Uhlenbeck process and
				generalizations: Particle dynamics under comb constraints and stochastic
				resetting},}\ }\href@noop {} {\bibfield  {journal} {\bibinfo  {journal}
			{Physical Review E}\ }\textbf {\bibinfo {volume} {107}},\ \bibinfo {pages}
		{054129} (\bibinfo {year} {2023})}\BibitemShut {NoStop}%
	\bibitem [{\citenamefont {Pal}, \citenamefont {Kundu},\ and\ \citenamefont
		{Evans}(2016)}]{pal2016diffusion}%
	\BibitemOpen
	\bibfield  {author} {\bibinfo {author} {\bibfnamefont {A.}~\bibnamefont
			{Pal}}, \bibinfo {author} {\bibfnamefont {A.}~\bibnamefont {Kundu}}, \ and\
		\bibinfo {author} {\bibfnamefont {M.~R.}\ \bibnamefont {Evans}},\ }\bibfield
	{title} {\enquote {\bibinfo {title} {Diffusion under time-dependent
				resetting},}\ }\href@noop {} {\bibfield  {journal} {\bibinfo  {journal}
			{Journal of Physics A: Mathematical and Theoretical}\ }\textbf {\bibinfo
			{volume} {49}},\ \bibinfo {pages} {225001} (\bibinfo {year}
		{2016})}\BibitemShut {NoStop}%
	\bibitem [{\citenamefont {Garc{\'\i}a-Valladares}\ \emph
		{et~al.}(2023)\citenamefont {Garc{\'\i}a-Valladares}, \citenamefont {Plata},
		\citenamefont {Prados},\ and\ \citenamefont {Manacorda}}]{garcia2023optimal}%
	\BibitemOpen
	\bibfield  {author} {\bibinfo {author} {\bibfnamefont {G.}~\bibnamefont
			{Garc{\'\i}a-Valladares}}, \bibinfo {author} {\bibfnamefont {C.~A.}\
			\bibnamefont {Plata}}, \bibinfo {author} {\bibfnamefont {A.}~\bibnamefont
			{Prados}}, \ and\ \bibinfo {author} {\bibfnamefont {A.}~\bibnamefont
			{Manacorda}},\ }\bibfield  {title} {\enquote {\bibinfo {title} {Optimal
				resetting strategies for search processes in heterogeneous environments},}\
	}\href@noop {} {\bibfield  {journal} {\bibinfo  {journal} {New Journal of
				Physics}\ }\textbf {\bibinfo {volume} {25}},\ \bibinfo {pages} {113031}
		(\bibinfo {year} {2023})}\BibitemShut {NoStop}%
	\bibitem [{\citenamefont {Rold{\'a}n}\ and\ \citenamefont
		{Gupta}(2017)}]{roldan2017path}%
	\BibitemOpen
	\bibfield  {author} {\bibinfo {author} {\bibfnamefont {{\'E}.}~\bibnamefont
			{Rold{\'a}n}}\ and\ \bibinfo {author} {\bibfnamefont {S.}~\bibnamefont
			{Gupta}},\ }\bibfield  {title} {\enquote {\bibinfo {title} {Path-integral
				formalism for stochastic resetting: Exactly solved examples and shortcuts to
				confinement},}\ }\href@noop {} {\bibfield  {journal} {\bibinfo  {journal}
			{Physical Review E}\ }\textbf {\bibinfo {volume} {96}},\ \bibinfo {pages}
		{022130} (\bibinfo {year} {2017})}\BibitemShut {NoStop}%
	\bibitem [{\citenamefont {Ku{\'s}mierz}\ and\ \citenamefont
		{Toyoizumi}(2019)}]{kusmierz2019robust}%
	\BibitemOpen
	\bibfield  {author} {\bibinfo {author} {\bibfnamefont {{\L}.}~\bibnamefont
			{Ku{\'s}mierz}}\ and\ \bibinfo {author} {\bibfnamefont {T.}~\bibnamefont
			{Toyoizumi}},\ }\bibfield  {title} {\enquote {\bibinfo {title} {Robust random
				search with scale-free stochastic resetting},}\ }\href@noop {} {\bibfield
		{journal} {\bibinfo  {journal} {Physical Review E}\ }\textbf {\bibinfo
			{volume} {100}},\ \bibinfo {pages} {032110} (\bibinfo {year}
		{2019})}\BibitemShut {NoStop}%
	\bibitem [{\citenamefont {Pinsky}(2020)}]{pinsky2020diffusive}%
	\BibitemOpen
	\bibfield  {author} {\bibinfo {author} {\bibfnamefont {R.~G.}\ \bibnamefont
			{Pinsky}},\ }\bibfield  {title} {\enquote {\bibinfo {title} {Diffusive search
				with spatially dependent resetting},}\ }\href@noop {} {\bibfield  {journal}
		{\bibinfo  {journal} {Stochastic Processes and their Applications}\ }\textbf
		{\bibinfo {volume} {130}},\ \bibinfo {pages} {2954--2973} (\bibinfo {year}
		{2020})}\BibitemShut {NoStop}%
	\bibitem [{\citenamefont {Sandev}\ \emph {et~al.}(2021)\citenamefont {Sandev},
		\citenamefont {Domazetoski}, \citenamefont {Iomin},\ and\ \citenamefont
		{Kocarev}}]{sandev2021diffusion}%
	\BibitemOpen
	\bibfield  {author} {\bibinfo {author} {\bibfnamefont {T.}~\bibnamefont
			{Sandev}}, \bibinfo {author} {\bibfnamefont {V.}~\bibnamefont {Domazetoski}},
		\bibinfo {author} {\bibfnamefont {A.}~\bibnamefont {Iomin}}, \ and\ \bibinfo
		{author} {\bibfnamefont {L.}~\bibnamefont {Kocarev}},\ }\bibfield  {title}
	{\enquote {\bibinfo {title} {Diffusion--advection equations on a comb:
				Resetting and random search},}\ }\href@noop {} {\bibfield  {journal}
		{\bibinfo  {journal} {Mathematics}\ }\textbf {\bibinfo {volume} {9}},\
		\bibinfo {pages} {221} (\bibinfo {year} {2021})}\BibitemShut {NoStop}%
	\bibitem [{\citenamefont {Pierce}(2022)}]{pierce2022advection}%
	\BibitemOpen
	\bibfield  {author} {\bibinfo {author} {\bibfnamefont {J.~K.}\ \bibnamefont
			{Pierce}},\ }\bibfield  {title} {\enquote {\bibinfo {title} {An
				advection-diffusion process with proportional resetting},}\ }\href@noop {}
	{\bibfield  {journal} {\bibinfo  {journal} {arXiv preprint arXiv:2204.07215}\
		} (\bibinfo {year} {2022})}\BibitemShut {NoStop}%
	\bibitem [{\citenamefont {Ray}, \citenamefont {Mondal},\ and\ \citenamefont
		{Reuveni}(2019)}]{ray2019peclet}%
	\BibitemOpen
	\bibfield  {author} {\bibinfo {author} {\bibfnamefont {S.}~\bibnamefont
			{Ray}}, \bibinfo {author} {\bibfnamefont {D.}~\bibnamefont {Mondal}}, \ and\
		\bibinfo {author} {\bibfnamefont {S.}~\bibnamefont {Reuveni}},\ }\bibfield
	{title} {\enquote {\bibinfo {title} {P{\'e}clet number governs transition to
				acceleratory restart in drift-diffusion},}\ }\href@noop {} {\bibfield
		{journal} {\bibinfo  {journal} {Journal of Physics A: Mathematical and
				Theoretical}\ }\textbf {\bibinfo {volume} {52}},\ \bibinfo {pages} {255002}
		(\bibinfo {year} {2019})}\BibitemShut {NoStop}%
	\bibitem [{\citenamefont {Stanislavsky}\ and\ \citenamefont
		{Weron}(2022)}]{stanislavsky2022subdiffusive}%
	\BibitemOpen
	\bibfield  {author} {\bibinfo {author} {\bibfnamefont {A.~A.}\ \bibnamefont
			{Stanislavsky}}\ and\ \bibinfo {author} {\bibfnamefont {A.}~\bibnamefont
			{Weron}},\ }\bibfield  {title} {\enquote {\bibinfo {title} {Subdiffusive
				search with home returns via stochastic resetting: a subordination scheme
				approach},}\ }\href@noop {} {\bibfield  {journal} {\bibinfo  {journal}
			{Journal of Physics A: Mathematical and Theoretical}\ }\textbf {\bibinfo
			{volume} {55}},\ \bibinfo {pages} {074004} (\bibinfo {year}
		{2022})}\BibitemShut {NoStop}%
	\bibitem [{\citenamefont {Ahmad}\ \emph {et~al.}(2019)\citenamefont {Ahmad},
		\citenamefont {Nayak}, \citenamefont {Bansal}, \citenamefont {Nandi},\ and\
		\citenamefont {Das}}]{ahmad2019first}%
	\BibitemOpen
	\bibfield  {author} {\bibinfo {author} {\bibfnamefont {S.}~\bibnamefont
			{Ahmad}}, \bibinfo {author} {\bibfnamefont {I.}~\bibnamefont {Nayak}},
		\bibinfo {author} {\bibfnamefont {A.}~\bibnamefont {Bansal}}, \bibinfo
		{author} {\bibfnamefont {A.}~\bibnamefont {Nandi}}, \ and\ \bibinfo {author}
		{\bibfnamefont {D.}~\bibnamefont {Das}},\ }\bibfield  {title} {\enquote
		{\bibinfo {title} {First passage of a particle in a potential under
				stochastic resetting: A vanishing transition of optimal resetting rate},}\
	}\href@noop {} {\bibfield  {journal} {\bibinfo  {journal} {Physical Review
				E}\ }\textbf {\bibinfo {volume} {99}},\ \bibinfo {pages} {022130} (\bibinfo
		{year} {2019})}\BibitemShut {NoStop}%
	\bibitem [{\citenamefont {Vinod}\ \emph {et~al.}(2022)\citenamefont {Vinod},
		\citenamefont {Cherstvy}, \citenamefont {Metzler},\ and\ \citenamefont
		{Sokolov}}]{vinod2022time}%
	\BibitemOpen
	\bibfield  {author} {\bibinfo {author} {\bibfnamefont {D.}~\bibnamefont
			{Vinod}}, \bibinfo {author} {\bibfnamefont {A.~G.}\ \bibnamefont {Cherstvy}},
		\bibinfo {author} {\bibfnamefont {R.}~\bibnamefont {Metzler}}, \ and\
		\bibinfo {author} {\bibfnamefont {I.~M.}\ \bibnamefont {Sokolov}},\
	}\bibfield  {title} {\enquote {\bibinfo {title} {Time-averaging and
				nonergodicity of reset geometric Brownian motion with drift},}\ }\href@noop
	{} {\bibfield  {journal} {\bibinfo  {journal} {Physical Review E}\ }\textbf
		{\bibinfo {volume} {106}},\ \bibinfo {pages} {034137} (\bibinfo {year}
		{2022})}\BibitemShut {NoStop}%
	\bibitem [{\citenamefont {Sarkar}\ and\ \citenamefont
		{Gupta}(2022)}]{sarkar2022biased}%
	\BibitemOpen
	\bibfield  {author} {\bibinfo {author} {\bibfnamefont {M.}~\bibnamefont
			{Sarkar}}\ and\ \bibinfo {author} {\bibfnamefont {S.}~\bibnamefont {Gupta}},\
	}\bibfield  {title} {\enquote {\bibinfo {title} {Biased random walk on random
				networks in presence of stochastic resetting: Exact results},}\ }\href@noop
	{} {\bibfield  {journal} {\bibinfo  {journal} {Journal of Physics A:
				Mathematical and Theoretical}\ }\textbf {\bibinfo {volume} {55}},\ \bibinfo
		{pages} {42LT01} (\bibinfo {year} {2022})}\BibitemShut {NoStop}%
	\bibitem [{\citenamefont {Montero}, \citenamefont {Mas{\'o}-Puigdellosas},\
		and\ \citenamefont {Villarroel}(2017)}]{montero2017continuous}%
	\BibitemOpen
	\bibfield  {author} {\bibinfo {author} {\bibfnamefont {M.}~\bibnamefont
			{Montero}}, \bibinfo {author} {\bibfnamefont {A.}~\bibnamefont
			{Mas{\'o}-Puigdellosas}}, \ and\ \bibinfo {author} {\bibfnamefont
			{J.}~\bibnamefont {Villarroel}},\ }\bibfield  {title} {\enquote {\bibinfo
			{title} {Continuous-time random walks with reset events: historical
				background and new perspectives},}\ }\href@noop {} {\bibfield  {journal}
		{\bibinfo  {journal} {The European Physical Journal B}\ }\textbf {\bibinfo
			{volume} {90}},\ \bibinfo {pages} {1--10} (\bibinfo {year}
		{2017})}\BibitemShut {NoStop}%
	\bibitem [{\citenamefont {Chechkin}\ \emph {et~al.}(2017)\citenamefont
		{Chechkin}, \citenamefont {Seno}, \citenamefont {Metzler},\ and\
		\citenamefont {Sokolov}}]{chechkin2017brownian}%
	\BibitemOpen
	\bibfield  {author} {\bibinfo {author} {\bibfnamefont {A.~V.}\ \bibnamefont
			{Chechkin}}, \bibinfo {author} {\bibfnamefont {F.}~\bibnamefont {Seno}},
		\bibinfo {author} {\bibfnamefont {R.}~\bibnamefont {Metzler}}, \ and\
		\bibinfo {author} {\bibfnamefont {I.~M.}\ \bibnamefont {Sokolov}},\
	}\bibfield  {title} {\enquote {\bibinfo {title} {Brownian yet non-Gaussian
				diffusion: from superstatistics to subordination of diffusing
				diffusivities},}\ }\href@noop {} {\bibfield  {journal} {\bibinfo  {journal}
			{Physical Review X}\ }\textbf {\bibinfo {volume} {7}},\ \bibinfo {pages}
		{021002} (\bibinfo {year} {2017})}\BibitemShut {NoStop}%
	\bibitem [{\citenamefont {Sposini}\ \emph {et~al.}(2018)\citenamefont
		{Sposini}, \citenamefont {Chechkin}, \citenamefont {Seno}, \citenamefont
		{Pagnini},\ and\ \citenamefont {Metzler}}]{sposini2018random}%
	\BibitemOpen
	\bibfield  {author} {\bibinfo {author} {\bibfnamefont {V.}~\bibnamefont
			{Sposini}}, \bibinfo {author} {\bibfnamefont {A.~V.}\ \bibnamefont
			{Chechkin}}, \bibinfo {author} {\bibfnamefont {F.}~\bibnamefont {Seno}},
		\bibinfo {author} {\bibfnamefont {G.}~\bibnamefont {Pagnini}}, \ and\
		\bibinfo {author} {\bibfnamefont {R.}~\bibnamefont {Metzler}},\ }\bibfield
	{title} {\enquote {\bibinfo {title} {Random diffusivity from stochastic
				equations: comparison of two models for Brownian yet non-Gaussian
				diffusion},}\ }\href@noop {} {\bibfield  {journal} {\bibinfo  {journal} {New
				Journal of Physics}\ }\textbf {\bibinfo {volume} {20}},\ \bibinfo {pages}
		{043044} (\bibinfo {year} {2018})}\BibitemShut {NoStop}%
	\bibitem [{\citenamefont {Postnikov}, \citenamefont {Chechkin},\ and\
		\citenamefont {Sokolov}(2020)}]{postnikov2020brownian}%
	\BibitemOpen
	\bibfield  {author} {\bibinfo {author} {\bibfnamefont {E.~B.}\ \bibnamefont
			{Postnikov}}, \bibinfo {author} {\bibfnamefont {A.}~\bibnamefont {Chechkin}},
		\ and\ \bibinfo {author} {\bibfnamefont {I.~M.}\ \bibnamefont {Sokolov}},\
	}\bibfield  {title} {\enquote {\bibinfo {title} {Brownian yet non-Gaussian
				diffusion in heterogeneous media: from superstatistics to homogenization},}\
	}\href@noop {} {\bibfield  {journal} {\bibinfo  {journal} {New Journal of
				Physics}\ }\textbf {\bibinfo {volume} {22}},\ \bibinfo {pages} {063046}
		(\bibinfo {year} {2020})}\BibitemShut {NoStop}%
	\bibitem [{\citenamefont {Elrick}(1962)}]{elrick1962source}%
	\BibitemOpen
	\bibfield  {author} {\bibinfo {author} {\bibfnamefont {D.}~\bibnamefont
			{Elrick}},\ }\bibfield  {title} {\enquote {\bibinfo {title} {Source functions
				for diffusion in uniform shear flow},}\ }\href@noop {} {\bibfield  {journal}
		{\bibinfo  {journal} {Australian Journal of Physics}\ }\textbf {\bibinfo
			{volume} {15}},\ \bibinfo {pages} {283--288} (\bibinfo {year}
		{1962})}\BibitemShut {NoStop}%
	\bibitem [{\citenamefont {Evans}\ and\ \citenamefont
		{Majumdar}(2014)}]{evans2014diffusion}%
	\BibitemOpen
	\bibfield  {author} {\bibinfo {author} {\bibfnamefont {M.~R.}\ \bibnamefont
			{Evans}}\ and\ \bibinfo {author} {\bibfnamefont {S.~N.}\ \bibnamefont
			{Majumdar}},\ }\bibfield  {title} {\enquote {\bibinfo {title} {Diffusion with
				resetting in arbitrary spatial dimension},}\ }\href@noop {} {\bibfield
		{journal} {\bibinfo  {journal} {Journal of Physics A: Mathematical and
				Theoretical}\ }\textbf {\bibinfo {volume} {47}},\ \bibinfo {pages} {285001}
		(\bibinfo {year} {2014})}\BibitemShut {NoStop}%
	\bibitem [{\citenamefont {Fuchs}, \citenamefont {Goldt},\ and\ \citenamefont
		{Seifert}(2016)}]{fuchs2016stochastic}%
	\BibitemOpen
	\bibfield  {author} {\bibinfo {author} {\bibfnamefont {J.}~\bibnamefont
			{Fuchs}}, \bibinfo {author} {\bibfnamefont {S.}~\bibnamefont {Goldt}}, \ and\
		\bibinfo {author} {\bibfnamefont {U.}~\bibnamefont {Seifert}},\ }\bibfield
	{title} {\enquote {\bibinfo {title} {Stochastic thermodynamics of
				resetting},}\ }\href@noop {} {\bibfield  {journal} {\bibinfo  {journal} {
				Europhysics Letters}\ }\textbf {\bibinfo {volume} {113}},\ \bibinfo {pages}
		{60009} (\bibinfo {year} {2016})}\BibitemShut {NoStop}%
	\bibitem [{\citenamefont {Olsen}\ \emph {et~al.}(2024)\citenamefont {Olsen},
		\citenamefont {Gupta}, \citenamefont {Mori},\ and\ \citenamefont
		{Krishnamurthy}}]{olsen_thermodynamic}%
	\BibitemOpen
	\bibfield  {author} {\bibinfo {author} {\bibfnamefont {K.~S.}\ \bibnamefont
			{Olsen}}, \bibinfo {author} {\bibfnamefont {D.}~\bibnamefont {Gupta}},
		\bibinfo {author} {\bibfnamefont {F.}~\bibnamefont {Mori}}, \ and\ \bibinfo
		{author} {\bibfnamefont {S.}~\bibnamefont {Krishnamurthy}},\ }\bibfield
	{title} {\enquote {\bibinfo {title} {Thermodynamic cost of finite-time
				stochastic resetting},}\ }\href@noop {} {\bibfield  {journal} {\bibinfo
			{journal} {Physical Review Research}\ }\textbf {\bibinfo {volume} {6}},\
		\bibinfo {pages} {033343} (\bibinfo {year} {2024})}\BibitemShut {NoStop}%
	\bibitem [{\citenamefont {Olsen}\ and\ \citenamefont
		{Gupta}(2024)}]{olsen2024thermodynamic}%
	\BibitemOpen
	\bibfield  {author} {\bibinfo {author} {\bibfnamefont {K.~S.}\ \bibnamefont
			{Olsen}}\ and\ \bibinfo {author} {\bibfnamefont {D.}~\bibnamefont {Gupta}},\
	}\bibfield  {title} {\enquote {\bibinfo {title} {Thermodynamic work of
				partial resetting},}\ }\href@noop {} {\bibfield  {journal} {\bibinfo
			{journal} {Journal of Physics A: Mathematical and Theoretical}\ }\textbf
		{\bibinfo {volume} {57}},\ \bibinfo {pages} {245001} (\bibinfo {year}
		{2024})}\BibitemShut {NoStop}%
	\bibitem [{\citenamefont {Mori}, \citenamefont {Olsen},\ and\ \citenamefont
		{Krishnamurthy}(2023)}]{mori2023entropy}%
	\BibitemOpen
	\bibfield  {author} {\bibinfo {author} {\bibfnamefont {F.}~\bibnamefont
			{Mori}}, \bibinfo {author} {\bibfnamefont {K.~S.}\ \bibnamefont {Olsen}}, \
		and\ \bibinfo {author} {\bibfnamefont {S.}~\bibnamefont {Krishnamurthy}},\
	}\bibfield  {title} {\enquote {\bibinfo {title} {Entropy production of
				resetting processes},}\ }\href@noop {} {\bibfield  {journal} {\bibinfo
			{journal} {Physical Review Research}\ }\textbf {\bibinfo {volume} {5}},\
		\bibinfo {pages} {023103} (\bibinfo {year} {2023})}\BibitemShut {NoStop}%
	\bibitem [{\citenamefont {Gupta}\ and\ \citenamefont
		{Plata}(2022)}]{Deepak2022_work}%
	\BibitemOpen
	\bibfield  {author} {\bibinfo {author} {\bibfnamefont {D.}~\bibnamefont
			{Gupta}}\ and\ \bibinfo {author} {\bibfnamefont {C.~A.}\ \bibnamefont
			{Plata}},\ }\bibfield  {title} {\enquote {\bibinfo {title} {Work fluctuations
				for diffusion dynamics submitted to stochastic return},}\ }\href@noop {}
	{\bibfield  {journal} {\bibinfo  {journal} {New Journal of Physics}\ } \textbf {\bibinfo {volume} {24}},\
		\bibinfo {pages} {113034} (\bibinfo {year} {2022})}\BibitemShut {NoStop}%
	\bibitem [{\citenamefont {Gupta}, \citenamefont {Plata},\ and\ \citenamefont
		{Pal}(2020)}]{gupta2020work}%
	\BibitemOpen
	\bibfield  {author} {\bibinfo {author} {\bibfnamefont {D.}~\bibnamefont
			{Gupta}}, \bibinfo {author} {\bibfnamefont {C.~A.}\ \bibnamefont {Plata}}, \
		and\ \bibinfo {author} {\bibfnamefont {A.}~\bibnamefont {Pal}},\ }\bibfield
	{title} {\enquote {\bibinfo {title} {Work fluctuations and Jarzynski equality
				in stochastic resetting},}\ }\href@noop {} {\bibfield  {journal} {\bibinfo
			{journal} {Physical Review Letters}\ }\textbf {\bibinfo {volume} {124}},\
		\bibinfo {pages} {110608} (\bibinfo {year} {2020})}\BibitemShut {NoStop}%
	\bibitem [{\citenamefont {Busiello}, \citenamefont {Gupta},\ and\ \citenamefont
		{Maritan}(2020)}]{Busiello2020uni}%
	\BibitemOpen
	\bibfield  {author} {\bibinfo {author} {\bibfnamefont {D.~M.}\ \bibnamefont
			{Busiello}}, \bibinfo {author} {\bibfnamefont {D.}~\bibnamefont {Gupta}}, \
		and\ \bibinfo {author} {\bibfnamefont {A.}~\bibnamefont {Maritan}},\
	}\bibfield  {title} {\enquote {\bibinfo {title} {Entropy production in
				systems with unidirectional transitions},}\ }\href {\doibase
		10.1103/PhysRevResearch.2.023011} {\bibfield  {journal} {\bibinfo  {journal}
			{Physical Review Research}\ }\textbf {\bibinfo {volume} {2}},\ \bibinfo {pages}
		{023011} (\bibinfo {year} {2020})}\BibitemShut {NoStop}%
	\bibitem [{\citenamefont {Pal}\ \emph {et~al.}(2023)\citenamefont {Pal},
		\citenamefont {Pal}, \citenamefont {Park},\ and\ \citenamefont
		{Lee}}]{pal2023thermodynamic}%
	\BibitemOpen
	\bibfield  {author} {\bibinfo {author} {\bibfnamefont {P.~S.}\ \bibnamefont
			{Pal}}, \bibinfo {author} {\bibfnamefont {A.}~\bibnamefont {Pal}}, \bibinfo
		{author} {\bibfnamefont {H.}~\bibnamefont {Park}}, \ and\ \bibinfo {author}
		{\bibfnamefont {J.~S.}\ \bibnamefont {Lee}},\ }\bibfield  {title} {\enquote
		{\bibinfo {title} {Thermodynamic trade-off relation for first passage time in
				resetting processes},}\ }\href@noop {} {\bibfield  {journal} {\bibinfo
			{journal} {Physical Review E}\ }\textbf {\bibinfo {volume} {108}},\ \bibinfo
		{pages} {044117} (\bibinfo {year} {2023})}\BibitemShut {NoStop}%
	\bibitem [{\citenamefont {Sunil}\ \emph {et~al.}(2023)\citenamefont {Sunil},
		\citenamefont {Blythe}, \citenamefont {Evans},\ and\ \citenamefont
		{Majumdar}}]{sunil2023cost}%
	\BibitemOpen
	\bibfield  {author} {\bibinfo {author} {\bibfnamefont {J.~C.}\ \bibnamefont
			{Sunil}}, \bibinfo {author} {\bibfnamefont {R.~A.}\ \bibnamefont {Blythe}},
		\bibinfo {author} {\bibfnamefont {M.~R.}\ \bibnamefont {Evans}}, \ and\
		\bibinfo {author} {\bibfnamefont {S.~N.}\ \bibnamefont {Majumdar}},\
	}\bibfield  {title} {\enquote {\bibinfo {title} {The cost of stochastic
				resetting},}\ }\href@noop {} {\bibfield  {journal} {\bibinfo  {journal}
			{Journal of Physics A: Mathematical and Theoretical}\ }\textbf {\bibinfo
			{volume} {56}},\ \bibinfo {pages} {395001} (\bibinfo {year}
		{2023})}\BibitemShut {NoStop}%
	\bibitem [{\citenamefont {Roberts}, \citenamefont {Sezik},\ and\ \citenamefont
		{Lardet}(2024)}]{Roberts_2024}%
	\BibitemOpen
	\bibfield  {author} {\bibinfo {author} {\bibfnamefont {C.}~\bibnamefont
			{Roberts}}, \bibinfo {author} {\bibfnamefont {E.}~\bibnamefont {Sezik}}, \
		and\ \bibinfo {author} {\bibfnamefont {E.}~\bibnamefont {Lardet}},\
	}\bibfield  {title} {\enquote {\bibinfo {title} {Ratchet-mediated resetting:
				current, efficiency, and exact solution},}\ }\href {\doibase
		10.1088/1751-8121/ad62c9} {\bibfield  {journal} {\bibinfo  {journal} {Journal
				of Physics A: Mathematical and Theoretical}\ }\textbf {\bibinfo {volume}
			{57}},\ \bibinfo {pages} {325001} (\bibinfo {year} {2024})}\BibitemShut
	{NoStop}%
	\bibitem [{\citenamefont {Goerlich}, \citenamefont {Keidar},\ and\
		\citenamefont {Roichman}(2024)}]{goerlich2024resetting}%
	\BibitemOpen
	\bibfield  {author} {\bibinfo {author} {\bibfnamefont {R.}~\bibnamefont
			{Goerlich}}, \bibinfo {author} {\bibfnamefont {T.~D.}\ \bibnamefont
			{Keidar}}, \ and\ \bibinfo {author} {\bibfnamefont {Y.}~\bibnamefont
			{Roichman}},\ }\bibfield  {title} {\enquote {\bibinfo {title} {Resetting as a
				swift equilibration protocol in an anharmonic potential},}\ }\href@noop {}
	{\bibfield  {journal} {\bibinfo  {journal} {Physical Review Research}\
		}\textbf {\bibinfo {volume} {6}},\ \bibinfo {pages} {033162} (\bibinfo {year}
		{2024})}\BibitemShut {NoStop}%
	\bibitem [{\citenamefont {Singh}(2024)}]{singh2024cost}%
	\BibitemOpen
	\bibfield  {author} {\bibinfo {author} {\bibfnamefont {P.}~\bibnamefont
			{Singh}},\ }\bibfield  {title} {\enquote {\bibinfo {title} {Cost-time
				trade-off in diffusion with stochastic return: optimal resetting potential
				and pareto front},}\ }\href@noop {} {\bibfield  {journal} {\bibinfo
			{journal} {arXiv preprint arXiv:2407.02071}\ } (\bibinfo {year}
		{2024})}\BibitemShut {NoStop}%
	\bibitem [{\citenamefont {Sunil}\ \emph {et~al.}(2024)\citenamefont {Sunil},
		\citenamefont {Blythe}, \citenamefont {Evans},\ and\ \citenamefont
		{Majumdar}}]{sunil2024minimizing}%
	\BibitemOpen
	\bibfield  {author} {\bibinfo {author} {\bibfnamefont {J.~C.}\ \bibnamefont
			{Sunil}}, \bibinfo {author} {\bibfnamefont {R.~A.}\ \bibnamefont {Blythe}},
		\bibinfo {author} {\bibfnamefont {M.~R.}\ \bibnamefont {Evans}}, \ and\
		\bibinfo {author} {\bibfnamefont {S.~N.}\ \bibnamefont {Majumdar}},\
	}\bibfield  {title} {\enquote {\bibinfo {title} {Minimizing the profligacy of
				searches with reset},}\ }\href@noop {} {\bibfield  {journal} {\bibinfo
			{journal} {arXiv preprint arXiv:2404.00215}\ } (\bibinfo {year}
		{2024})}\BibitemShut {NoStop}%
	\bibitem [{\citenamefont {Tal-Friedman}\ \emph {et~al.}(2024)\citenamefont
		{Tal-Friedman}, \citenamefont {Keidar}, \citenamefont {Reuveni},\ and\
		\citenamefont {Roichman}}]{tal2024smart}%
	\BibitemOpen
	\bibfield  {author} {\bibinfo {author} {\bibfnamefont {O.}~\bibnamefont
			{Tal-Friedman}}, \bibinfo {author} {\bibfnamefont {T.~D.}\ \bibnamefont
			{Keidar}}, \bibinfo {author} {\bibfnamefont {S.}~\bibnamefont {Reuveni}}, \
		and\ \bibinfo {author} {\bibfnamefont {Y.}~\bibnamefont {Roichman}},\
	}\bibfield  {title} {\enquote {\bibinfo {title} {Smart resetting: An
				energy-efficient strategy for stochastic search processes},}\ }\href@noop {}
	{\bibfield  {journal} {\bibinfo  {journal} {arXiv preprint arXiv:2409.10108}\
		} (\bibinfo {year} {2024})}\BibitemShut {NoStop}%
	\bibitem [{\citenamefont {Van~den Broeck}, \citenamefont {Sancho},\ and\
		\citenamefont {San~Miguel}(1982)}]{van1982harmonically}%
	\BibitemOpen
	\bibfield  {author} {\bibinfo {author} {\bibfnamefont {C.}~\bibnamefont
			{Van~den Broeck}}, \bibinfo {author} {\bibfnamefont {J.}~\bibnamefont
			{Sancho}}, \ and\ \bibinfo {author} {\bibfnamefont {M.}~\bibnamefont
			{San~Miguel}},\ }\bibfield  {title} {\enquote {\bibinfo {title} {Harmonically
				bound Brownian motion in flowing fluids},}\ }\href@noop {} {\bibfield
		{journal} {\bibinfo  {journal} {Physica A: Statistical Mechanics and its
				Applications}\ }\textbf {\bibinfo {volume} {116}},\ \bibinfo {pages}
		{448--461} (\bibinfo {year} {1982})}\BibitemShut {NoStop}%
	\bibitem [{\citenamefont {Besga}\ \emph {et~al.}(2020)\citenamefont {Besga},
		\citenamefont {Bovon}, \citenamefont {Petrosyan}, \citenamefont {Majumdar},\
		and\ \citenamefont {Ciliberto}}]{besga2020optimal}%
	\BibitemOpen
	\bibfield  {author} {\bibinfo {author} {\bibfnamefont {B.}~\bibnamefont
			{Besga}}, \bibinfo {author} {\bibfnamefont {A.}~\bibnamefont {Bovon}},
		\bibinfo {author} {\bibfnamefont {A.}~\bibnamefont {Petrosyan}}, \bibinfo
		{author} {\bibfnamefont {S.~N.}\ \bibnamefont {Majumdar}}, \ and\ \bibinfo
		{author} {\bibfnamefont {S.}~\bibnamefont {Ciliberto}},\ }\bibfield  {title}
	{\enquote {\bibinfo {title} {Optimal mean first-passage time for a Brownian
				searcher subjected to resetting: experimental and theoretical results},}\
	}\href@noop {} {\bibfield  {journal} {\bibinfo  {journal} {Physical Review
				Research}\ }\textbf {\bibinfo {volume} {2}},\ \bibinfo {pages} {032029}
		(\bibinfo {year} {2020})}\BibitemShut {NoStop}%
	\bibitem [{\citenamefont {Tal-Friedman}\ \emph {et~al.}(2020)\citenamefont
		{Tal-Friedman}, \citenamefont {Pal}, \citenamefont {Sekhon}, \citenamefont
		{Reuveni},\ and\ \citenamefont {Roichman}}]{friedman2020exp}%
	\BibitemOpen
	\bibfield  {author} {\bibinfo {author} {\bibfnamefont {O.}~\bibnamefont
			{Tal-Friedman}}, \bibinfo {author} {\bibfnamefont {A.}~\bibnamefont {Pal}},
		\bibinfo {author} {\bibfnamefont {A.}~\bibnamefont {Sekhon}}, \bibinfo
		{author} {\bibfnamefont {S.}~\bibnamefont {Reuveni}}, \ and\ \bibinfo
		{author} {\bibfnamefont {Y.}~\bibnamefont {Roichman}},\ }\bibfield  {title}
	{\enquote {\bibinfo {title} {Experimental realization of diffusion with
				stochastic resetting},}\ }\href@noop {} {\bibfield  {journal} {\bibinfo
			{journal} {J. Phys. Chem. Lett.}\ }\textbf {\bibinfo {volume} {11}},\
		\bibinfo {pages} {7350--7355} (\bibinfo {year} {2020})}\BibitemShut {NoStop}%
	\bibitem [{\citenamefont {Goerlich}\ \emph {et~al.}(2023)\citenamefont
		{Goerlich}, \citenamefont {Li}, \citenamefont {Pires}, \citenamefont
		{Hervieux}, \citenamefont {Manfredi},\ and\ \citenamefont
		{Genet}}]{goerlich2023experimental}%
	\BibitemOpen
	\bibfield  {author} {\bibinfo {author} {\bibfnamefont {R.}~\bibnamefont
			{Goerlich}}, \bibinfo {author} {\bibfnamefont {M.}~\bibnamefont {Li}},
		\bibinfo {author} {\bibfnamefont {L.~B.}\ \bibnamefont {Pires}}, \bibinfo
		{author} {\bibfnamefont {P.-A.}\ \bibnamefont {Hervieux}}, \bibinfo {author}
		{\bibfnamefont {G.}~\bibnamefont {Manfredi}}, \ and\ \bibinfo {author}
		{\bibfnamefont {C.}~\bibnamefont {Genet}},\ }\bibfield  {title} {\enquote
		{\bibinfo {title} {Experimental test of Landauer's principle for stochastic
				resetting},}\ }\href@noop {} {\bibfield  {journal} {\bibinfo  {journal}
			{arXiv preprint arXiv:2306.09503}\ } (\bibinfo {year} {2023})}\BibitemShut
	{NoStop}%
\end{thebibliography}
\end{document}